\documentclass[a4paper,8pt]{article}
\pdfoutput=1
\usepackage{jcappub} 
\usepackage[T1]{fontenc} 
\usepackage{soul}
\usepackage{titlesec}
\titleformat{\paragraph}
{\normalfont\normalsize\bfseries}{\theparagraph}{1em}{}
\titlespacing*{\paragraph}
{0pt}{3.25ex plus 1ex minus .2ex}{1.5ex plus .2ex}
\usepackage{pifont}
\usepackage{lscape}         
\usepackage{graphics}    
\usepackage{hyperref}      
\usepackage{color}  
\usepackage{caption}
\usepackage{subcaption}    
\usepackage{ifthen}                 \usepackage{orcidlink}

\usepackage{amssymb}
\usepackage{mathrsfs}
\usepackage{bbold}
\usepackage{float}                     
\usepackage{multirow}
\usepackage{tikz}
\usepackage{mathtools}%
\usepackage{cellspace,tabularx}%
\usepackage{comment}
\usepackage{xcolor}
\usepackage{soul}
\def\nn{\nonumber}
\def\ds{{\rm d}s}

\def\da{{\rm d}a}

\def\be{\begin{equation}}
\def\ee{\end{equation}}
\def\barr{\begin{array}{lr}}
\def\earr{\end{array}}
\def\bea{\begin{eqnarray}}
\def\eea{\end{eqnarray}}
\def\nn{\nonumber}

\def\k{\kappa}

\begin{document}
\title{Understanding the non-Gaussian nature of Galactic foreground emissions towards small scales}

\author[a,b,c]{Fazlu Rahman\orcidlink{0000-0002-6292-1855},}
\author[c,d]{Pravabati Chingangbam\orcidlink{0000-0002-7385-8273},}
\author[a,b]{Kevin Huffenberger\orcidlink{0000-0001-7109-0099},}
\author[e,f]{Tuhin Ghosh\orcidlink{0000-0001-6088-3034},}
\author[g] {Dmitri Pogosyan\orcidlink{0000-0002-7998-6823},}
\author[h,i]{Tarun Souradeep\orcidlink{0000-0003-3764-8102},}
\author[d]{Changbom Park\orcidlink{0000-0001-9521-6397}}
\affiliation[a]{Mitchell Institute for Fundamental Physics and Astronomy, Texas A\&M University, College Station, TX, 77843, USA}
\affiliation[b]{Department of Physics and Astronomy, Texas A\&M University, College Station, TX, 77843, USA}
\affiliation[c]{Indian Institute of Astrophysics, Koramangala II Block, Bengaluru,  560 034, India}
\affiliation[d]{Korea Institute for Advanced Study (KIAS), 85 Hoegiro, Dongdaemun-gu, Seoul, 02455, Republic of Korea}
\affiliation[e]{School for Physical Sciences, National Institute of Science Education and Research, HBNI Jatni-752050, India}
\affiliation[f]{Homi Bhabha National Institute, Training School Complex, Anushakti Nagar, Mumbai 400094, India}
\affiliation[g]{Department of Physics, University of Alberta, Edmonton, AB T6G 2E1, Canada}
\affiliation[h]{Raman Research Institute, C. V. Raman Avenue, Sadashivanagar, Bengaluru, 560 080, India}
\affiliation[i]{Inter University Centre for Astronomy and Astrophysics, Post Bag 4, Ganeshkhind, Pune, 411 007, India}
\emailAdd{fazlu@tamu.edu}
\emailAdd{prava@iiap.res.in}
\emailAdd{khuffenberger@tamu.edu}
\emailAdd{tghosh@niser.ac.in}
\emailAdd{pogosyan@ualberta.ca}
\emailAdd{tarun@rri.res.in}
\emailAdd{cbp@kias.re.kr}

\abstract{
We present a unified, multi-scale study of non-Gaussianity of Galactic foreground emissions using Minkowski Functionals and generalized skewness–kurtosis parameters,  
focusing on the characterization of small-scale non-Gaussianity and its underlying physical origin. 
We find that all foreground components studied exhibit a remarkably universal non-Gaussian nature dominated by excess kurtosis, whose shape remains stable across angular scales despite large differences in emission physics. Focusing on thermal dust, we perform a detailed comparison between observed maps (GNILC and Planck 545 GHz) and dust model realizations (PySM and filament-based models) to assess the performance of state-of-the-art models in reproducing the observed non-Gaussian properties. At the global level, GNILC and PySM display closely matched kurtosis behavior over the angular scales where the GNILC reconstruction is reliable, while the filament-based model produces substantially weaker skewness and kurtosis signals. For PySM, however, a patch-based analysis reveals statistically significant regional variations, indicating that while the model reproduces the overall non-Gaussian amplitude and scale dependence, it does not fully capture the spatial variability of the observed kurtosis signal. 
Using simple PDF-based toy models, we demonstrate that the universal kurtosis signature arises from the combination of heavy-tailed one-point statistics and steep large-scale spatial correlations, while its detailed amplitude and scale dependence depend on the underlying foreground physics.
These results identify excess kurtosis as a robust statistical fingerprint of Galactic foregrounds and provide a practical framework for validating small-scale foreground models for future CMB analyses.}

\maketitle
\section{Introduction}

\noindent
Adequate mitigation of contamination by Galactic foreground emissions 
remains one of the primary challenges to achieving the science goals of upcoming CMB experiments. As CMB instruments reach higher sensitivity and finer angular resolution, even subtle imperfections in foreground modelling can propagate into biases in component-separated maps, lensing reconstructions, and searches for primordial B-modes~\cite{Remazeilles:2015hpa,Beck:2020dhe,Hertig:2024adq,Bianchini:2025dhf,SimonsObservatory:2025avm}. The situation is further complicated by the fact that Galactic emissions arise from multiple physical processes -- synchrotron radiation, free–free emission, Anomalous Microwave Emission (AME), thermal dust and so on, each with distinct frequency behaviour, spatial variability, and correlations within the interstellar medium. These complexities make it difficult for simple models to capture the full structure of real sky maps and motivate the development of statistical tools capable of probing their morphology and higher-order features.   

Current observational data of Galactic emissions remain limited in both frequency coverage and angular resolution. For example, thermal dust maps observed at high frequencies, where the dust contribution dominates, must be extrapolated using a modified blackbody spectrum with spectral index~$\beta$. Even small uncertainties in the spatial variation of~$\beta$ can induce non-Gaussian residuals and lead to significant biases in the inferred tensor-to-scalar ratio $r$~\cite{Liu:2025hmw_SOcomplexity}. To model foreground structure on angular scales beyond the resolution of current templates, several approaches have been developed~\cite{Thorne2017,Remazeilles:2014mba}. The Python Sky Model (PySM), widely used for CMB forecasting, generates non-Gaussian small-scale dust and synchrotron fluctuations through its \texttt{logpoltens} prescription~\cite{PySM2025}. Filament-inspired models~\cite{Hervias-Caimapo:DustFilaments} construct small-scale dust fluctuations based on the anisotropic morphology of observed filaments in the dust maps. More recently, machine-learning–based approaches have emerged that learn large-scale statistical properties of the sky and generate high-resolution non-Gaussian realizations consistent with them~\cite{Krachmalnicoff:ForSE,Melis:2023,Yao:2024xkg_Forse+}. Accurately modelling small-scale non-Gaussian structure in foregrounds is particularly important, as it has been shown that such non-Gaussianity can bias CMB lensing reconstruction and, consequently, impact searches for primordial B-modes~\cite{Abril-Cabezas:2025whd}. Yet its physical origin, universality, and scale-dependence across different Galactic emissions are not fully established, motivating further investigation of their statistical properties.

Historically, Galactic emissions have been characterized primarily through power spectrum analyses~\cite{Burigana2006,MivilleDeschenes:dustCl,Lazarian2012,Mertsch:2013pua,Planck:2018dust,Martire2022,2024:Stalpes}. Since these emissions are not Gaussian fields, the power spectrum alone is insufficient and higher order statistics are required to fully understand their nature. The three-point function, bispectrum, and skewness–kurtosis parameters have also been used to characterize non-Gaussianity in thermal dust and synchrotron emission~\cite{Ben-David:2015b,Jung2018,Rana:2018oft,Coulton:2019bnz,von1019}. Scattering transforms provide an alternative approach to quantify higher-order correlations in these maps~\cite{Allys:2019fna,Saydjari:2021,Delouis:2022yyt}. 
Complementary to these, Minkowski Functionals (MFs) provide a complete set of morphological descriptors for random fields and have emerged as powerful real-space tools for characterizing non-Gaussianity~\cite{Adler:1981,Tomita:1986,Mecke:1994,Gott:1990,Matsubara:1995ns,Schmalzing:1997,Naselski:1998,Dolgov:1999,Matsubara:2003,Pogosyan:2011,Gay:2012}. 
MFs capture  information about the topology and geometry of structures. Unlike harmonic-space statistics, which are most sensitive to correlations at specific multipoles, MFs respond directly to phase correlations and higher-order interactions, making them well suited for detecting non-Gaussian features generated by nonlinear astrophysical processes. 
In cosmology, MFs have been widely applied across a broad range of problems. 
In particular for the CMB, they have been used to search for primordial non-Gaussianity, test statistical isotropy, characterize lensing-induced mode coupling, identify secondary anisotropies, and probe signatures in polarization fields~\cite{Schmalzing:1998,Chingangbam:2009,Hikage:2012,Ganesan:2014lqa,Chingangbam:2017sap,Planck:2019evm,Carones:2022rbv,Bashir:2025gen,Sabyr:2024kcu,Chen:2026jlr}.  

Early studies used topological measures, such as the genus statistic, to probe foreground contamination in CMB data~\cite{Park:2001,Park:2002}. Building on this, MFs were used to detect residual foreground contamination in foreground subtracted CMB data from WMAP~\cite{Chingangbam:2013}. Since then, they have become important tools in the study of Galactic foregrounds, capturing signatures of magnetized turbulence, filamentary dust structures, synchrotron fluctuations, and other interstellar medium (ISM)-driven non-Gaussian features~\cite{Rana:2018oft,Seta:2020,Rahman:2021azv,Rahman:2022edq,Martire:2023ytg,Kalberla:2023}. More recently, MFs and related morphological descriptors have been applied to validate synthetic sky models and to probe residual foreground contamination, where conventional power-spectrum-based diagnostics often prove insufficient~\cite{Krachmalnicoff:ForSE,Hervias-Caimapo:DustFilaments,PySM2025,Ranucci:2025lao,Puglisi:2026fbj}. 
This broad range of applications highlights the versatility of MFs as robust real-space statistics, capable of isolating subtle non-Gaussian signatures arising from both cosmological signals and complex Galactic processes. 
In~\cite{Rahman:2021azv} (hereafter RCG2021), MFs applied to the Haslam 408 MHz map revealed that Galactic synchrotron emission exhibits a kurtosis-dominated non-Gaussian signature whose shape remains approximately stable across angular scales, while its amplitude varies with scale. This behaviour, characterized by excess kurtosis and subdominant skewness, was later confirmed in Planck temperature and polarization maps~\cite{Rahman:2022edq,Martire:2023ytg}. These results indicate a persistent non-Gaussian component not captured by power-spectrum analyses, likely linked to turbulent ISM structure.

In this work, we broaden the morphology-based non-Gaussianity analysis originally developed for synchrotron emission and extend it systematically across all major Planck temperature foregrounds, including free–free, AME, and thermal dust emission. We analyze the statistical properties of the observed maps obtained via multiple component separation methods, including the GNILC and \textit{Commander} products~\cite{Planck_2015_X,Planck:2016_GNILC}, and compare them with commonly used foreground realizations such as PySM~\cite{PySM2025} and the filament-based \texttt{DUSTFILAMENTS} model~\cite{Hervias-Caimapo:DustFilaments}. This comparison allows us to assess whether commonly used foreground models reproduce the scale-dependent non-Gaussian signatures observed in real-sky maps. 
A central result of this work is that the same kurtosis-dominated non-Gaussian trend previously seen in synchrotron emission also appears across all major Galactic foregrounds at small angular scales, despite their very different physical origins. To better understand the origin of this common behavior, we perform controlled experiments using random-field toy models with known one-point probability distribution functions (PDFs). These tests allow us to investigate how the observed MF signatures depend on both the one-point PDF and the spatial correlations of the field.

The remainder of this paper is organized as follows. Section~\ref{sec:DataSets_Preprocessing} describes the data sets, pre-processing steps, and the statistical framework used in this work, including Minkowski Functionals and skewness-kurtosis parameters. Section~\ref{sec:results_Galactic} presents the main results on the morphology and scale-dependent non-Gaussianity of Galactic emissions, along with comparisons between Planck products and dust model realizations. In Section~\ref{sec:OriginofKurt}, we discuss the origin of the small-scale kurtosis-dominated non-Gaussianity making use of controlled toy models. We summarize our findings and discuss their implications for foreground modelling and future CMB analyses in Section~\ref{sec:DiscussionConclusion}.

\section{Data sets and analysis methodology}
\label{sec:DataSets_Preprocessing}

This section describes the foreground data sets used in this work, along with the masking and filtering procedures applied prior to the non-Gaussianity analysis. The analysis methodology using MFs and SK parameters is then described.

\subsection{Foreground maps studied}
\label{sec:FgDataSets}

We study a combination of observed and simulated Galactic foreground maps that are commonly used in CMB foreground studies. These include component-separated products from the Planck mission, and simulated dust realizations designed to represent small-scale foreground structures. Together, these data sets allow us to compare non-Gaussian signatures across different emission mechanisms, resolutions, and modeling assumptions.

We begin with the Planck \textit{Commander} component-separated foreground products, which provide full-sky maps of the major diffuse Galactic emission components, including synchrotron, free-free, AME, and thermal dust. These maps are derived through a Bayesian parametric fitting approach applied to multi-frequency Planck maps, and are provided at a common HEALPix\footnote{\url{https://healpix.sourceforge.io}} resolution of $N_{\rm side}=256$ with an effective FWHM of $60$ arcmin~\cite{Planck:X}. While limited in angular resolution, the \textit{Commander} products offer a consistent baseline for comparing the statistical properties of non-Gaussianity across different foreground components.

Next, for a detailed investigation of thermal dust non-Gaussianity at small angular scales, we use high-resolution dust intensity maps from Planck. Our main observational data sets are the Planck GNILC thermal dust map~\cite{Planck:2016_GNILC} and the Planck $545$ GHz frequency map of intensity~\cite{Planck_2018III_HFI}. The GNILC map provides a component-separated estimate of diffuse dust emission using a scale-dependent needlet-based approach designed to suppress CIB and instrumental noise. Since the native GNILC map has a spatially varying effective beam, we use the beam-homogenized version with a uniform angular resolution of 21.8 arcmin provided within the PySM framework~\cite{PySM2025}, constructed by combining the first six needlet scales. This ensures a well-defined and spatially uniform beam for our scale-dependent analysis. The 545 GHz map, on the other hand, is a high signal-to-noise, dust dominated frequency map with minimal contribution from other Galactic components. Its native angular resolution is approximately $4.8$ arcmin. 
We use this map as a frequency-based reference for comparison with the component-separated dust maps.

To complement the observational analysis and to assess how well current models reproduce the observed non-Gaussian properties of dust emission, we also analyze dust model realizations. We consider two classes of models that are widely used in CMB analyses: the PySM dust model~\cite{PySM2025} and the \texttt{DUSTFILAMENTS} model~\cite{Hervias-Caimapo:DustFilaments} . While both are anchored to the large-scale dust morphology traced by the Planck GNILC dust map, they differ substantially in how small-scale structure is generated, providing a useful testbed for assessing the sensitivity of our statistics to different foreground modeling approaches. In PySM, we focus on the \texttt{d9} realization at 353\,GHz. In this model, small-scale dust structure is generated using the \texttt{logpoltens} formalism, which introduces non-Gaussian fluctuations while maintaining approximate consistency with the observed dust power spectrum. The \texttt{d9}\footnote{The closely related PySM \texttt{d10} model introduces frequency decorrelation through spatially varying small-scale $\beta$ realizations when scaling the dust template to frequencies other than 353\,GHz. Since \texttt{d9} and \texttt{d10} are identical at 353\,GHz, the results presented here apply equally to both models.} realization is widely used in forecasting and validation studies for current and upcoming CMB experiments, making it a natural benchmark for comparison with observational data. The \texttt{DUSTFILAMENTS} model incorporates explicitly modeled filamentary ISM structure and magnetic-field alignment, providing a physically motivated description of small-scale dust morphology while reproducing the observed power-spectrum characteristics of Galactic dust emission.
 
It should be noted that the maps are provided in the native units of their respective experiments. Since unit conversions correspond only to a global rescaling of the field and do not affect its morphology, topology, or non-Gaussian statistics considered here, we do not standardize the maps to a common unit system. All low-resolution \textit{Commander} foreground products are analyzed at their native angular resolution of $60$ arcmin, while all high resolution maps are analyzed at a common resolution of $21.8$ arcmin, corresponding to the effective beam of the {\it GNILC} dust map. Accordingly, the \textit{Commander} products are analyzed at $N_{\rm side}=256$, while the higher-resolution maps are downgraded to $N_{\rm side}=512$.
A summary of all foreground data products used in this work, along with their resolutions and units, is given in Table~\ref{tab:foreground_maps}.

\begin{table}[t]
\centering
\renewcommand{\arraystretch}{1.35}
\setlength{\tabcolsep}{8pt}

\begin{tabular}{l@{\hspace{20pt}}cccc}
\hline
\textbf{Map} & \textbf{FWHM} (arcmin) & \textbf{Units} & \textbf{Eff. Freq.} (GHz) & \textbf{Reference} \\
\hline
\multicolumn{5}{l}{\textit{\textit{Commander} Products}} \\
Synchrotron & 60 & K$_{\rm RJ}$ & 0.408 & \cite{Planck:X} \\
Free-free (EM) & " & cm$^{-6}$ pc & 30 & " \\
Thermal dust & " & K$_{\rm RJ}$ & 545 & " \\
AME & " & " & 22.8 & " \\
\hline
\multicolumn{5}{l}{\textit{High-Resolution Thermal Dust Maps}} \\
Planck GNILC (PySM) & 21.8 & MJy/sr & 353 & \cite{Planck:2016_GNILC,PySM2025} \\
Planck 545 GHz & " & " & 545 & \cite{Planck_2018III_HFI} \\
\hline
\multicolumn{5}{l}{\textit{Dust Model Realizations}} \\
PySM \texttt{d9} & " & MJy/sr & 353 & \cite{PySM2025} \\
\texttt{DUSTFILAMENTS} & " & $\mu {\rm K}_{\rm CMB}$ & " & \cite{Hervias-Caimapo:DustFilaments} \\
\hline
\end{tabular}
\caption{Summary of the observed and simulated foreground data sets analyzed in this work, including their angular resolution, units, effective reference frequency, and literature references. EM denotes the Emission Measure. The label ``GNILC (PySM)’’ refers to the beam-homogenized GNILC dust map provided within the PySM framework, which we use in this analysis.}
\label{tab:foreground_maps}
\end{table}


\subsection{Data processing and Gaussian-isotropic simulations}
\label{sec:maskingbpsims}

\begin{figure}[t]
    \centering
    \includegraphics[scale=1.20]{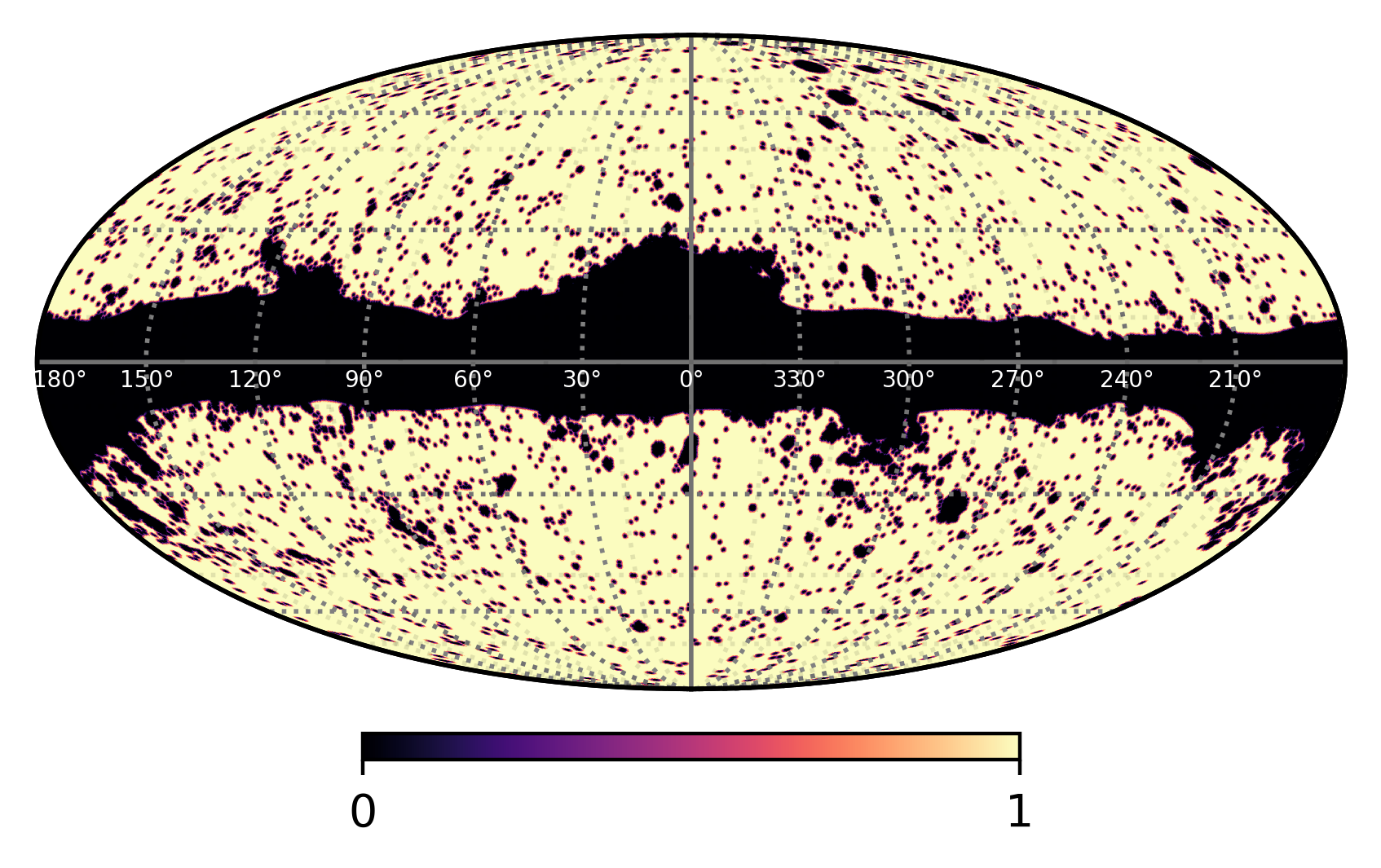}\\
    \caption{Final analysis mask used in this work, obtained after combining the \texttt{GalPlane80} and \texttt{CMB Common} masks. The mask is apodized using a \texttt{NaMaster C1} scheme with an apodization scale of $1^\circ$, retaining a sky fraction of 61\%.}
    \label{fig:Common_mask}
\end{figure}

To probe non-Gaussianity, all foreground maps are analyzed after applying a combination of sky masking and harmonic-space bandpass filtering. Masking is required to exclude regions dominated by bright Galactic emission and compact sources, while filtering is used to isolate fluctuations on well-defined angular scales. 

We begin by applying the Planck Galactic Plane mask (\texttt{GalPlane80}) that retains 80\% of the sky. This initial mask suppresses the brightest Galactic plane emission and is applied prior to bandpass filtering to mitigate ringing and leakage from bright pixels. Following RCG2021, we next apply a smooth hyperbolic-tangent (tanh) filter in harmonic space to isolate the angular scales of interest and study the small-scale structure of each map. The filter is defined as
$f_{\ell} =
(1/2)
\left[
1 +
\tanh\left(
\frac{\ell-\ell_{c}}{\Delta \ell}
\right)
\right]
\times
(1/2)
\left[
1 -
\tanh\left(
\frac{\ell-\ell^{*}}{\Delta \ell}
\right)
\right],$
where $\ell_{c}$ controls the progressive suppression of large-scale modes, while $\ell^{*}$ imposes an upper cutoff determined by the effective beam of the map.
For the $N_{\rm side}=256$ analyses, we set the upper cutoff to $\ell^{*}=180$, while for $N_{\rm side}=512$, we adopt $\ell^{*}=700$. We use a transition width of $\Delta \ell = 10$ for the filter and vary $\ell_c$ as one of the main parameters of the analysis. We also verified the robustness of our results against different harmonic tapers, including cosine, sigmoid, and Gaussian filters. In all cases, the results remain consistent at the sub-percent level, with no systematic differences in amplitude or scale dependence.

Next, we apply a final analysis mask defined as the union of the \texttt{GalPlane80} mask and the Planck \texttt{CMB Common} mask. The Common mask excludes both the Galactic plane and known compact sources identified by Planck, ensuring that residual contamination from bright point sources and strongly masked regions does not bias the non-Gaussianity measurements. Both the \texttt{GalPlane80} and \texttt{CMB Common} masks are apodized using a \texttt{C1} apodization scheme implemented in \texttt{NaMaster}\footnote{\url{https://namaster.readthedocs.io/}} with a scale of $1^{\circ}$ to minimize artifacts from sharp mask boundaries. To further suppress boundary-related leakage, we restrict the analysis to pixels with apodized mask values greater than 0.9. The resulting effective mask retains a sky fraction of approximately 61\% and is shown in Figure~\ref{fig:Common_mask}.
To test the sensitivity to the Galactic-plane cut, we repeat the analysis with \texttt{GalPlane60}, \texttt{GalPlane70}, and \texttt{GalPlane90}. The amplitudes of the measured non-Gaussianity vary mildly due to spatial inhomogeneity, but the scale dependence and overall trends reported in this work remain unchanged.

To quantify the non-Gaussian deviations, we generate Gaussian isotropic simulations for every map studied. For each case, we construct 1000 realizations using the angular power spectrum of the filtered and masked map, ensuring that the simulations reproduce the two-point statistics of the map under consideration.  The power spectra are estimated using \texttt{NaMaster}, which accounts for the effects of incomplete sky coverage. These simulations provide the Gaussian reference against which the non-Gaussian signatures reported in this work are quantified.

\subsection{Minkowski Functionals and skewness-kurtosis parameters}
\label{sec:MF_math}

For a random scalar field $u(\hat{n})$ on the unit sphere, excursion sets $Q_\nu$ are defined to be regions where $u \ge\nu\sigma$, 
with $\sigma$ being the  standard deviation of the field. $\nu$ is called the {\em threshold}. The morphology of excursion sets are then quantified by three scalar Minkowski Functionals (MFs), defined {\em per unit area} by, 
\be
V_0(\nu) \equiv \frac1A\int_{Q_{\nu}} \da, \quad
V_1(\nu) \equiv \frac{1}{4A}\int_{\partial Q_{\nu}} \ds, \quad
V_2(\nu) \equiv \frac{1}{2\pi A}\int_{\partial Q_{\nu}}  \k\,\ds,
\ee
where $A$ is the total unmasked area ($4\pi$ for full sky), $\da$ is the surface element, $\ds$ is the line element along the excursion boundary $\partial Q_\nu$, and $\kappa$ is the geodesic curvature~\cite{Tomita:1986,Mecke:1994,Schmalzing:1997}. $V_0$, $V_1$, and $V_2$ are the area fraction, boundary length, and Euler characteristic (or genus, ignoring a term proportional to the excursion area) of the excursion sets, respectively. 

The ensemble expectations  of MFs are determined by the nature of the random field encoded in the joint probability density function (JPDF)  of the field, and its first and  higher order derivatives. For a mildly non-Gaussian isotropic field with zero mean, these expectations can be expanded in terms of Hermite polynomials~\cite{Gay:2012,Matsubara:2011,Pogosyan:2009,Adler:1981,Tomita:1986} with coefficients given by a set of three generalized skewness and four generalized kurtosis cumulants. The details are given in Appendix~\ref{SecA:weaklynGMF}.
The skewness and kurtosis cumulants are defined in terms of the field $u$, its gradient $\nabla u$, and Laplacian $\nabla^{2} u$,  as follows,
\begin{eqnarray}
S_0 &=& \frac{\langle u^{3}\rangle_c}{\sigma^{4}}, \quad S_1=\frac{\langle u^{2}\nabla^{2}u\rangle_c}{\sigma^{2}\sigma^{2}_{1}},  \quad S_2= \frac{2\langle |\nabla u|^{2}\nabla^{2}u\rangle_c}{\sigma_{1}^{4}}, \label{eqn:skew}\\
K_0 &=& \frac{\langle u^{4}\rangle_c}{\sigma^{6}}, \ K_1=\frac{\langle u^{3}\nabla^{2}u\rangle_c}{\sigma^{4}\sigma^{2}_{1}}, \  
K_2=\frac{2\langle u|\nabla u|^{2}\nabla^{2}u\rangle_c+\langle |\nabla u|^{4}\rangle_c}{\sigma^{2}\sigma_{1}^{4}}, \ K_3=\frac{\langle |\nabla u|^{4}\rangle_c}{2\sigma^{2}\sigma_{1}^{4}}. \label{eqn:kurt}
\end{eqnarray}
In the above, $\sigma_1$ denotes the standard deviation of $\nabla u$. 
The subscript $c$ indicates that these quantities are the connected cumulants. Explicit relations between connected cumulants and ordinary moments can be found in Ref.~\cite{Matsubara:2020} and RCG2021. Skewness parameters generate contributions that are odd in $\nu$, whereas kurtosis parameters generate even contributions. This parity structure governs the symmetry of the non-Gaussian corrections to the MFs. 
Further, note that $S_0$ and $K_0$ are properties of the 1-point PDF of the field, while the other skewness and kurtosis parameters encode the full JPDF. For mildly non-Gaussian fields,  MFs can be characterised by the skewness and kurtosis parameters up to second order in $\sigma$. Since in this paper we examine each field in a multiscale manner by varying the bandpass filter scale $\ell_c$ from fully non-Gaussian to mildly non-Gaussian regimes, we carry out our analysis using both MFs and  generalized skewness and kurtosis parameters. Throughout this work, we present these parameters in the dimensionless combinations $S_i\sigma$ and $K_i\sigma^2$, which directly determine the leading non-Gaussian corrections to the MFs in the perturbative expansion.

We compute the MFs following the numerical {algorithm} of Schmalzing \& Górski~\cite{Schmalzing:1998}. 
To quantify non-Gaussianity,
we compare the measured MFs with Gaussian expectations obtained from 
Gaussian realizations constructed with the same angular power spectrum. Data processing steps including bandpass filtering and masking are identically applied to the data and Gaussian maps for a fair comparison. We define the MF deviations as
\be
\Delta V_k(\nu) =
\frac{V_k^{\rm data}(\nu) - \langle V_k^{\rm G}(\nu) \rangle}
{V_k^{\rm G,max}},
\ee
where $\langle V_k^{\rm G} \rangle$ denotes the ensemble average over Gaussian realizations and $V_k^{\rm G,max}$ is the maximum value of the corresponding Gaussian MF. In conjunction, we compute the generalized skewness and kurtosis parameters defined in Eqs.~\eqref{eqn:skew}--\eqref{eqn:kurt}, which quantify the leading cumulant contributions entering the weakly non-Gaussian expansion of the MFs. These parameters provide a complementary measure of higher-order correlations and help interpret the MF deviations in terms of skewness- and kurtosis-driven non-Gaussianity.

\section{Results – morphology and non-Gaussianity of Galactic emissions}
\label{sec:results_Galactic}

We now apply the formalism described above to Galactic foreground maps and examine their morphological properties as a function of angular scale. We begin with the Planck \textit{Commander} component-separated products, which provide all-sky reconstructions of the major diffuse foreground emissions. After establishing the main trends among these foregrounds, we focus on thermal dust emission in greater detail, comparing different Planck dust maps and dust model realizations to investigate the robustness of the observed non-Gaussian features.

\subsection{Planck \textit{Commander} foreground maps} \label{sec:planckcommander}

\begin{figure}[t]
    \centering    \includegraphics[width=\linewidth]{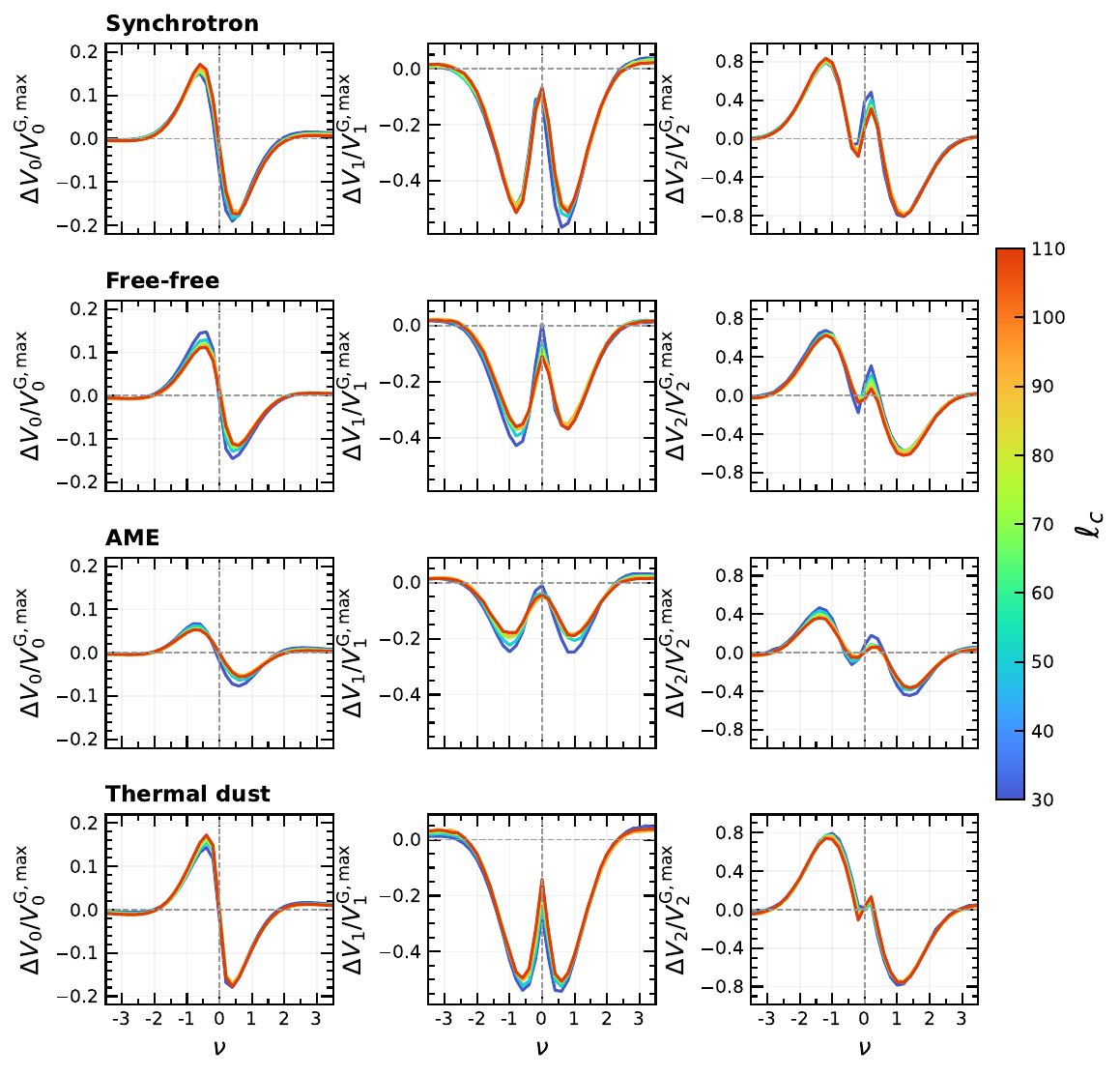}
    \caption{Minkowski Functional non-Gaussian deviations ($\Delta V_k$, with $k=0,1,2$ corresponding to the left, middle, and right columns, respectively) for the Planck \textit{Commander} foreground components, shown for different values of the lower multipole cut $\ell_c$. The upper multipole cut is fixed at $\ell^{*}=180$. The deviations exhibit a characteristic kurtosis-dominated morphology, while their amplitudes and scale-dependence differ among the foreground components.}
    \label{fig:fg_comm256}
\end{figure}

We analyze the Planck \textit{Commander} foreground maps for synchrotron, free–free, AME, and thermal dust emission. A key question is whether the scale-dependent kurtosis signature previously identified in synchrotron emission from Planck and Haslam maps also appears in other Galactic foreground components. This comparison also allows us to examine how the amplitude and scale dependence of the non-Gaussian signal vary among these foreground components, which arise from different physical processes and exhibit distinct spatial morphologies. Although synchrotron emission was studied in earlier work, we include it here for completeness and as a baseline for comparison.

We compute the three scalar MFs ($V_k$) and examine their deviations from the Gaussian expectation ($\Delta V_k$) as a function of angular scale. Figure~\ref{fig:fg_comm256} shows $\Delta V_k$ as a function of the threshold $\nu$ for the four foreground components and different values of $\ell_c$.
 
Across all \textit{Commander} foreground maps, we find clear non-Gaussian deviations relative to the Gaussian simulations. The amplitude of these deviations depends on angular scale, being generally stronger at large scales and weaker toward smaller scales. Despite this overall decrease in amplitude, the shape of the MF deviations remains remarkably similar across the different foreground components. Specifically, $\Delta V_0$ and $\Delta V_2$ exhibit predominantly antisymmetric deviations about $\nu=0$, while $\Delta V_1$ shows a more symmetric structure. These symmetry properties are consistent with the Hermite polynomial combinations expected from fourth-order cumulants (generalized kurtosis terms) in the perturbative MF expansion, rather than the skewness terms (see Appendix~\ref{SecA:weaklynGMF}). This pattern extends our previous findings for synchrotron emission in RCG2021 and~\cite{Rahman:2022edq}, indicating that a similar kurtosis-dominated statistical signature is present in foreground components with otherwise very different emission mechanisms. The amplitude of the signal, however, varies between components, with synchrotron and thermal dust showing the largest deviations, while free–free and especially AME exhibit comparatively weaker non-Gaussianity.

\begin{figure}[t]
    \centering
    \includegraphics[width=\linewidth]{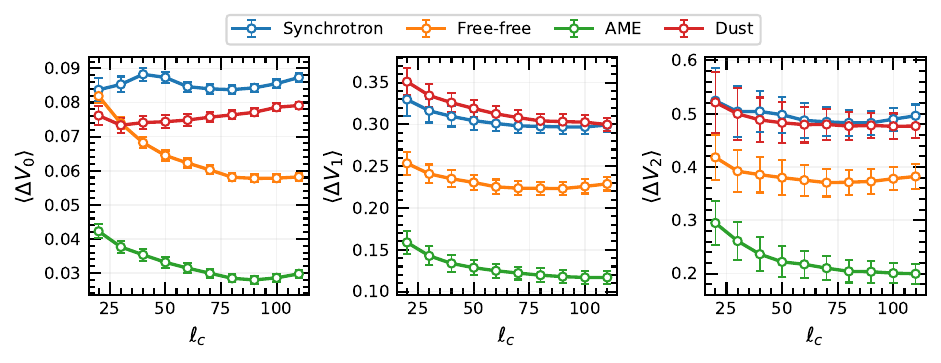}\\
    \caption{Average absolute MF non-Gaussian deviations ($\langle{\Delta V_k}\rangle$) as a function of bandpass scale $\ell_c$ for the Planck \textit{Commander} foreground components. Both the amplitude and scale dependence of the non-Gaussian deviations vary between foregrounds, with synchrotron and thermal dust showing the largest deviations over the range shown. Error bars denote the $2\sigma$ scatter.}
    \label{fig:comm256_dvKavg}
\end{figure}

To quantify the scale dependence of the non-Gaussian signal, we study how the MF deviations vary with $\ell_c$. We summarize their overall amplitude using the average absolute deviation of $\Delta V_k$,
\begin{equation}
\langle{\Delta V_k}\rangle=\frac{1}{N_{\rm tot}}\sum_{\nu=-2}^{2}|\Delta V_k(\nu)|,
\end{equation}
where $N_{\rm tot}=21$ corresponds to the number of threshold bins in the range $-2 \le \nu \le 2$ with spacing $\Delta\nu=0.2$\footnote{We ignore correlations between the thresholds $\nu$ in this definition.}.

Figure~\ref{fig:comm256_dvKavg} compares the scale dependence of $\langle{\Delta V_k}\rangle$ for the different Galactic foreground components. The amplitudes differ systematically between components: synchrotron and thermal dust show the largest deviations, free–free exhibits intermediate levels, while AME remains comparatively weaker. Although the same kurtosis-dominated signature is present in all foregrounds, its strength varies substantially between components. The scale dependence also differs modestly between emissions. In particular, we note a mild increase of $\langle{\Delta V_k}\rangle$ with $\ell_c$ for synchrotron, similar to that reported in our earlier analysis, where residual point-source contamination was identified as a possible contributor to the larger deviations at higher $\ell_c$~\cite{Rahman:2022edq}.

The persistence of the kurtosis-dominated signature over a broad range of angular scales, together with its appearance in foreground components with very different emission mechanisms, suggests that the observed non-Gaussianity is not tied to the details of a particular emission process. From a foreground-modelling perspective, this is encouraging: the presence of a shared higher-order statistical signature implies that similar statistical prescriptions may be applicable when constructing small-scale foreground realizations. These statistics also provide a useful benchmark for assessing the statistical fidelity of new models. At the same time, the emergence of a similar signature in these emissions hints at an underlying connection to the turbulent and intermittent nature of the interstellar medium.

To investigate this possibility further, we also examined a set of external tracers of the interstellar medium, including infrared maps such as IRAS $100\,\mu\mathrm{m}$ and WISE $12\,\mu\mathrm{m}$, as well as HI and H$\alpha$ emission~\cite{Miville-Desch:2005,Meisner:2014,HI4PI:2016,Finkbeiner:2003}. Despite probing different ISM phases and arising from distinct emission mechanisms, these tracers show qualitatively similar kurtosis-dominated MF signatures. The appearance of this pattern across such diverse tracers supports the interpretation that the observed kurtosis-driven non-Gaussianity is a general statistical property of ISM structure.

\subsection{Thermal dust emission}
\label{sec:ng_dustmaps}

Having established that kurtosis-dominated non-Gaussianity is a common feature of Galactic foreground emission, we now focus on thermal dust, the primary foreground for current and upcoming CMB experiments. The dust maps considered here also allow us to extend the analysis to substantially smaller angular scales than the low-resolution \textit{Commander} study discussed above. The availability of multiple observational tracers and model realizations makes dust an ideal test case for testing the robustness of the observed non-Gaussian signatures at these scales.

In this section, we analyze both observed dust maps, namely the Planck GNILC dust map and the Planck 545\,GHz intensity map, as well as dust realizations from PySM and \texttt{DUSTFILAMENTS}. This comparison provides a direct test of how well current dust models reproduce the observed scale-dependent statistical properties of Galactic dust emission.

\subsubsection{Observed dust tracers: Planck GNILC and 545 GHz maps}

\begin{figure}[t]
    \centering
    \includegraphics[width=\linewidth]{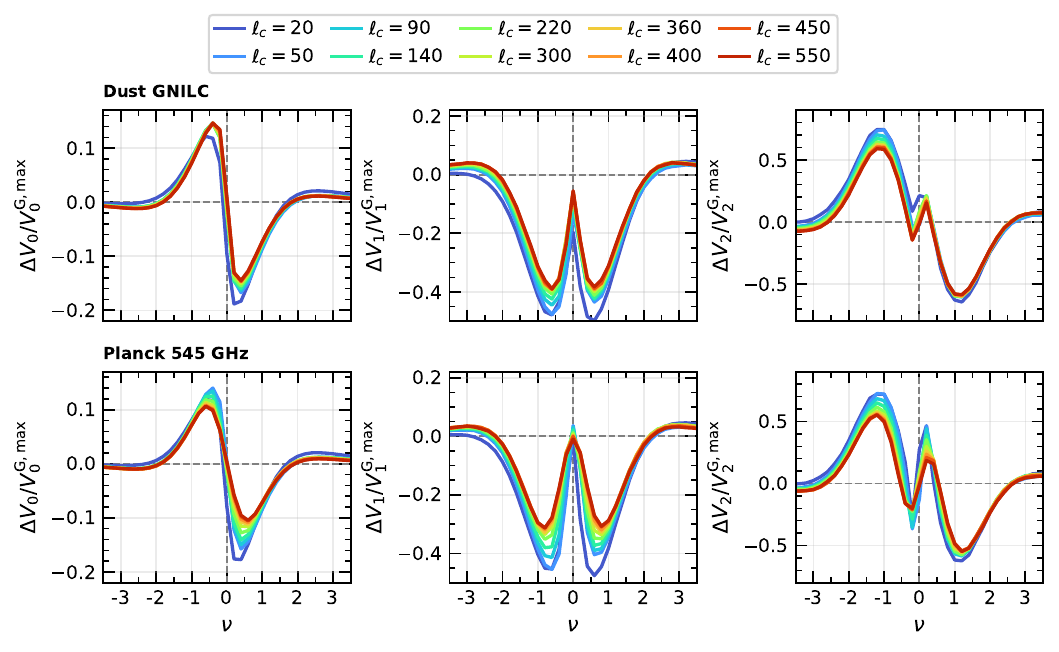}
    \includegraphics[width=\linewidth]{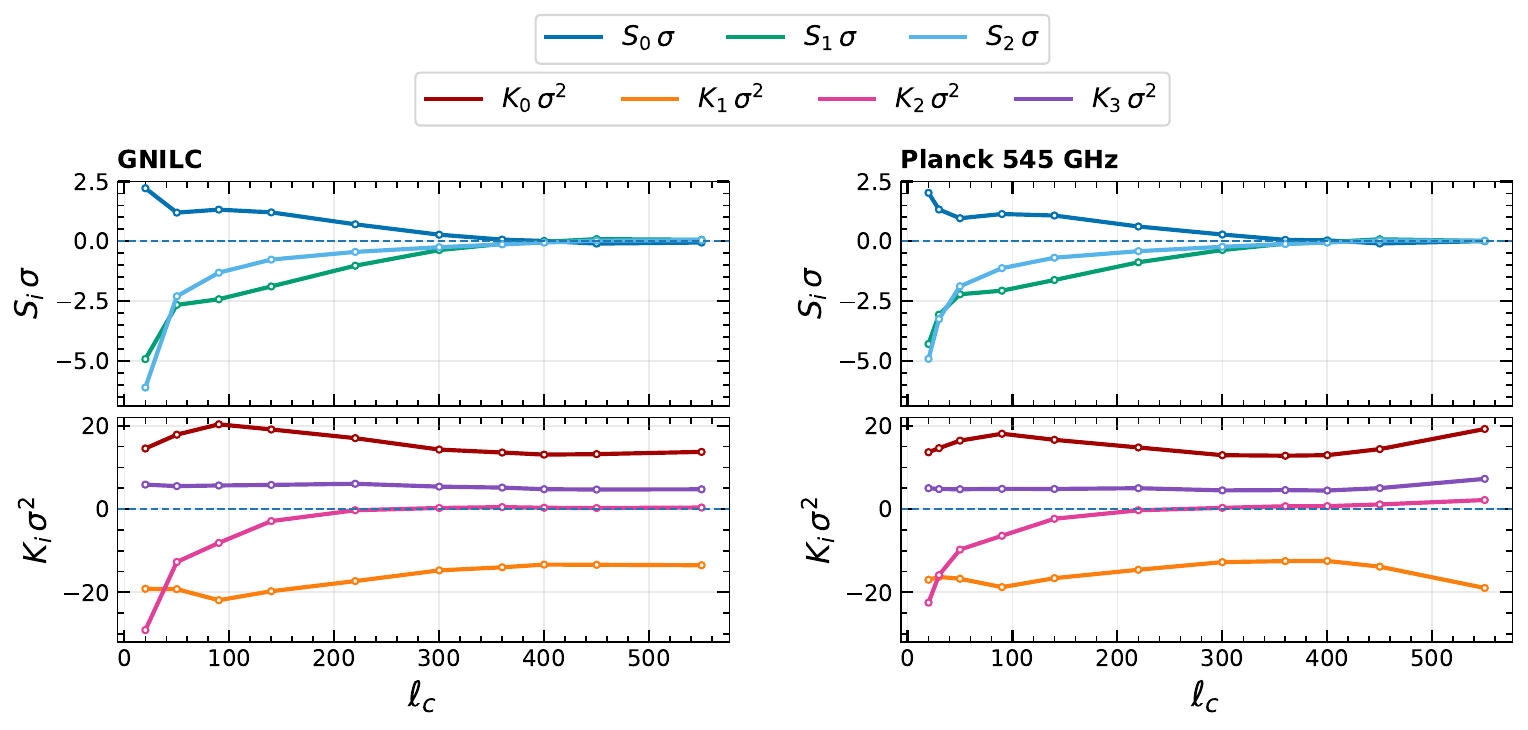}
    
    \caption{MF non-Gaussian deviations ($\Delta V_k$; top) and generalized skewness–kurtosis parameters (bottom) for the GNILC and Planck 545 GHz frequency maps.} 
    \label{fig:dV_thermdust}
\end{figure}

We begin our dust analysis with the Planck GNILC dust map and the Planck 545\,GHz intensity map. The former provides a component-separated estimate of thermal dust emission, while the latter serves as a direct observational tracer of dust morphology. The top panel of Figure~\ref{fig:dV_thermdust} shows the MF deviations as a function of threshold $\nu$ for maps filtered at different values of $\ell_c$. The bottom panel shows the corresponding SK parameters for the two maps. 

We find that both maps show clear non-Gaussian MF deviations across all three functionals. The shapes of $\Delta V_0$, $\Delta V_1$, and $\Delta V_2$ follow the kurtosis-driven MF shapes previously identified in the low-resolution \textit{Commander} foreground products, with amplitudes that gradually decrease toward smaller angular scales. The close agreement between the GNILC and 545\,GHz results indicates that these non-Gaussian features are robust properties of Galactic dust emission, independent of the particular dust tracer used.

SK parameters provide a more quantitative view of these observations. As a function of scale $\ell_c$, the skewness parameters decrease rapidly in amplitude and approach zero with increasing $\ell_c$. In contrast, the kurtosis parameters remain comparatively stable over the full range of $\ell_c$, making the non-Gaussianity increasingly kurtosis-dominated at high multipoles. For both the GNILC and 545\,GHz maps, the kurtosis-related terms exceed the skewness contributions at all scales considered here, providing direct evidence that the observed MF deviations are primarily associated with kurtosis rather than skewness. Interestingly, the scale dependence of the kurtosis parameters differs from that of the MF deviations themselves, whose amplitudes gradually decrease toward smaller angular scales. This illustrates the complementary nature of the two statistics: while the SK parameters capture specific combinations of three-point and four-point correlations of the field and its derivatives, the MF deviations provide a more complete description of the non-Gaussian morphology of the field.

RCG2021 found a similar hierarchy in Galactic synchrotron emission using the Haslam map, with the kurtosis parameters dominating over the skewness contributions across angular scales and giving rise to the kurtosis-dominated MF morphology. The recurrence of the same hierarchy in thermal dust suggests that the skewness-suppressed, kurtosis-dominated structure is a common feature of Galactic foreground emission, manifesting both in the MF morphology and in the underlying SK parameters.

For completeness, we repeated the analysis using the high-resolution \textit{Commander} thermal dust map~\cite{Planck_2015_X}. The resulting MF and SK measurements exhibit a similar scale dependence, with skewness remaining subdominant to kurtosis across the full range of scales considered here. The amplitudes are slightly smaller, possibly reflecting differences in the component-separation procedure. Those results are presented in Appendix~\ref{secA:commander_545}. In the same appendix, we also extend the Planck 545\,GHz analysis to $\ell^\ast=1600$, probing angular scales beyond the baseline choice of $\ell^\ast=700$.

\subsubsection{Dust models: PySM and \texttt{DUSTFILAMENTS}}
\label{sec:syntheticmaps}

\begin{figure}
    \centering    \includegraphics[width=\linewidth]{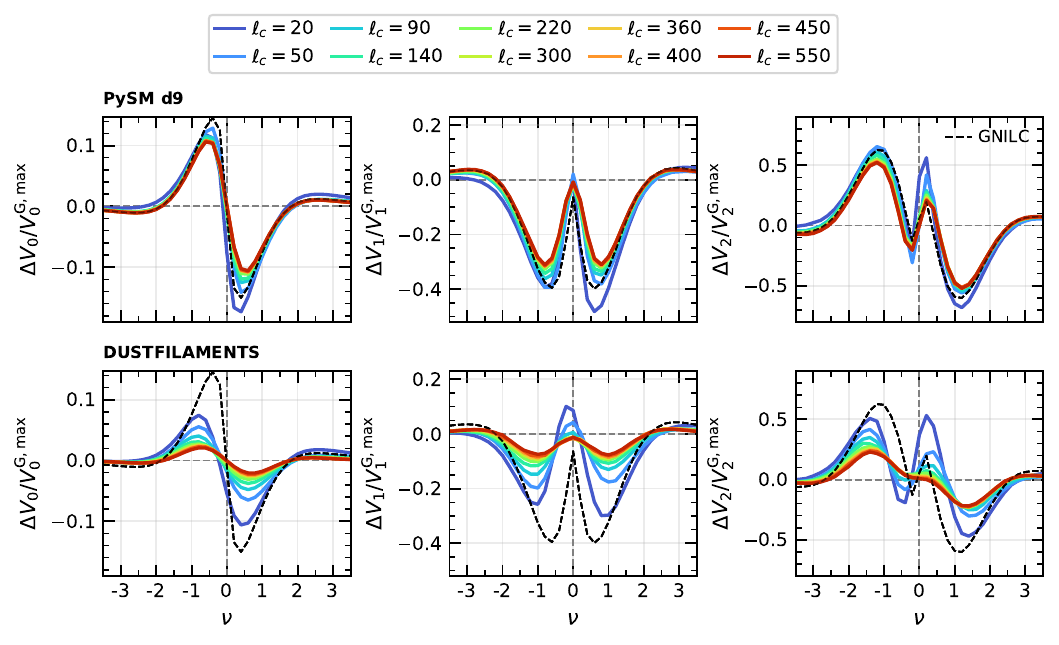} \\
    \includegraphics[width=\linewidth]{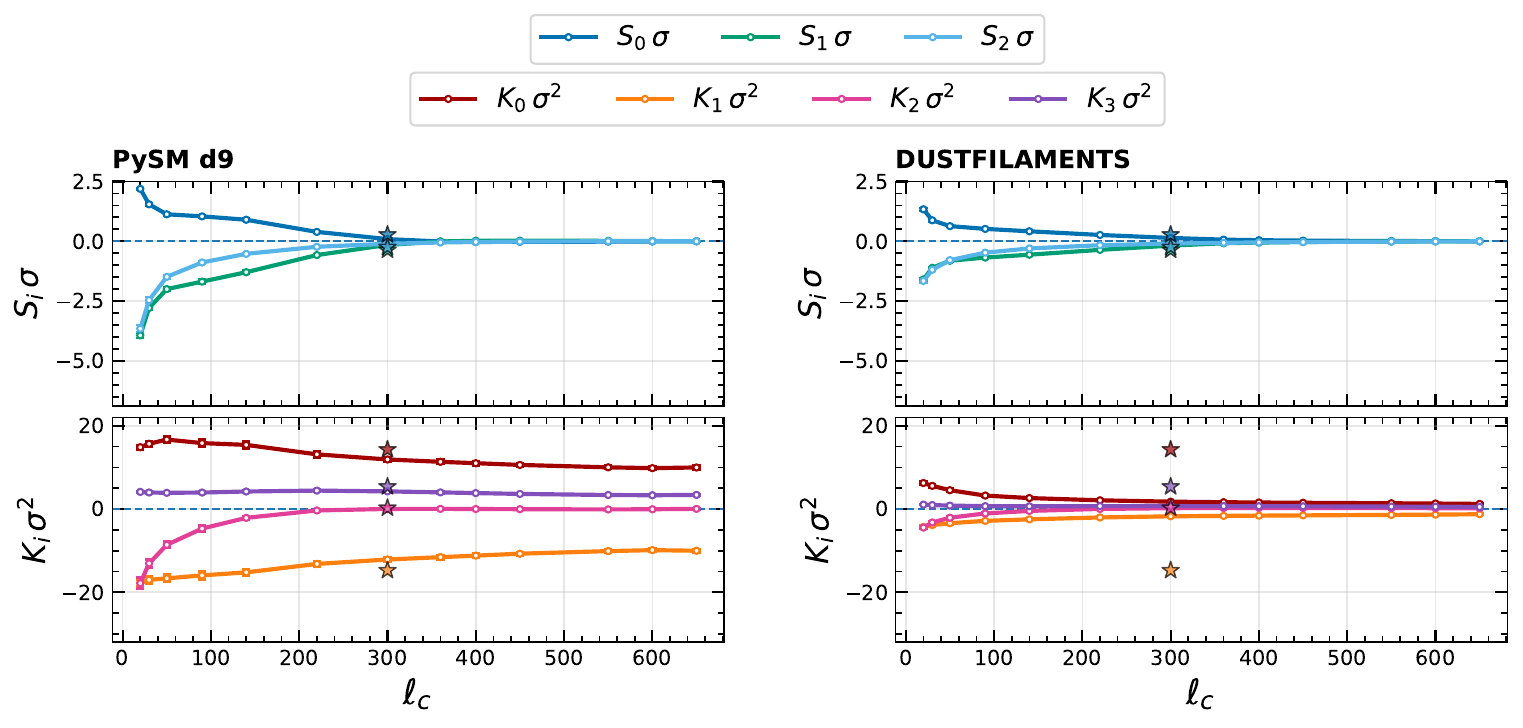}
    \caption{MF non-Gaussian deviations ($\Delta V_k$; top) and SK parameters (bottom) for the PySM \texttt{d9} and \texttt{DUSTFILAMENTS} thermal dust simulations. The dashed black curves in the MF panels show the GNILC reference at $\ell_c=300$, while the colored star markers in the SK panels denote the corresponding GNILC SK values at the same scale. The error bars in the SK parameter plots denote $2\sigma$ uncertainties, obtained using 100 realizations of the respective model.}
    \label{fig:dV_dustsims}
\end{figure}

We now turn to model realizations of thermal dust and examine how well they reproduce the scale-dependent non-Gaussian features seen in the data. Here, we analyze the PySM \texttt{d9} and filament-based \texttt{DUSTFILAMENTS} dust models at 353\,GHz and compare them directly with the observed dust maps. Figure~\ref{fig:dV_dustsims} shows the MF deviations and corresponding SK parameters for the two dust models. For PySM, the results shown correspond to the mean MF deviations and SK parameters obtained from 100 independent realizations of the \texttt{d9} model (\texttt{d11} in the PySM nomenclature). The \texttt{DUSTFILAMENTS} results are computed in the same way from 100 independent realizations.

The PySM dust model reproduces the kurtosis-dominated MF morphology seen in the observational maps. The MF deviations closely resemble those of the GNILC and Planck 545\,GHz maps, with amplitudes that decrease with increasing $\ell_c$ and approach an approximately scale-independent regime above the injection scale of the PySM small-scale realizations ($\ell \sim 100$). The SK parameters show the same qualitative behaviour: kurtosis parameters remain dominant over the full range of scales, while the skewness parameters rapidly decrease toward zero at high multipoles. In short, the PySM \texttt{logpoltens} construction captures both the skewness–kurtosis balance and the scale dependence of the non-Gaussianity observed in Galactic dust.

The \texttt{DUSTFILAMENTS} model shows a different trend. On large angular scales, the MF deviations broadly resemble those seen in the observational maps, but the amplitudes of both the MF deviations and SK parameters are substantially smaller. Toward smaller angular scales, the non-Gaussian signal decreases rapidly and the kurtosis-dominated MF morphology becomes progressively weaker. This is reflected in the SK parameters, where the skewness and kurtosis terms become increasingly comparable with increasing $\ell_c$, in contrast to the strongly kurtosis-dominated regime seen in the observed maps and reproduced by PySM. Overall, the current \texttt{DUSTFILAMENTS} realization underestimates the level of non-Gaussianity present in the Planck dust maps. This is consistent with the conclusions of~\cite{Hervias-Caimapo:DustFilaments}, which found that the model does not fully reproduce the observed small-scale non-Gaussianity in Planck maps. One possible reason for this discrepancy is that the current implementation is primarily tuned to match the observed dust power spectrum, which by itself does not strongly constrain higher-order statistics. The observed SK parameters may therefore provide additional constraints for refining the filament model and improving its agreement with the data.

\subsubsection{GNILC vs PySM d9: patch-based comparisons}

\begin{figure}
    \centering
    \includegraphics[width=0.98\linewidth]{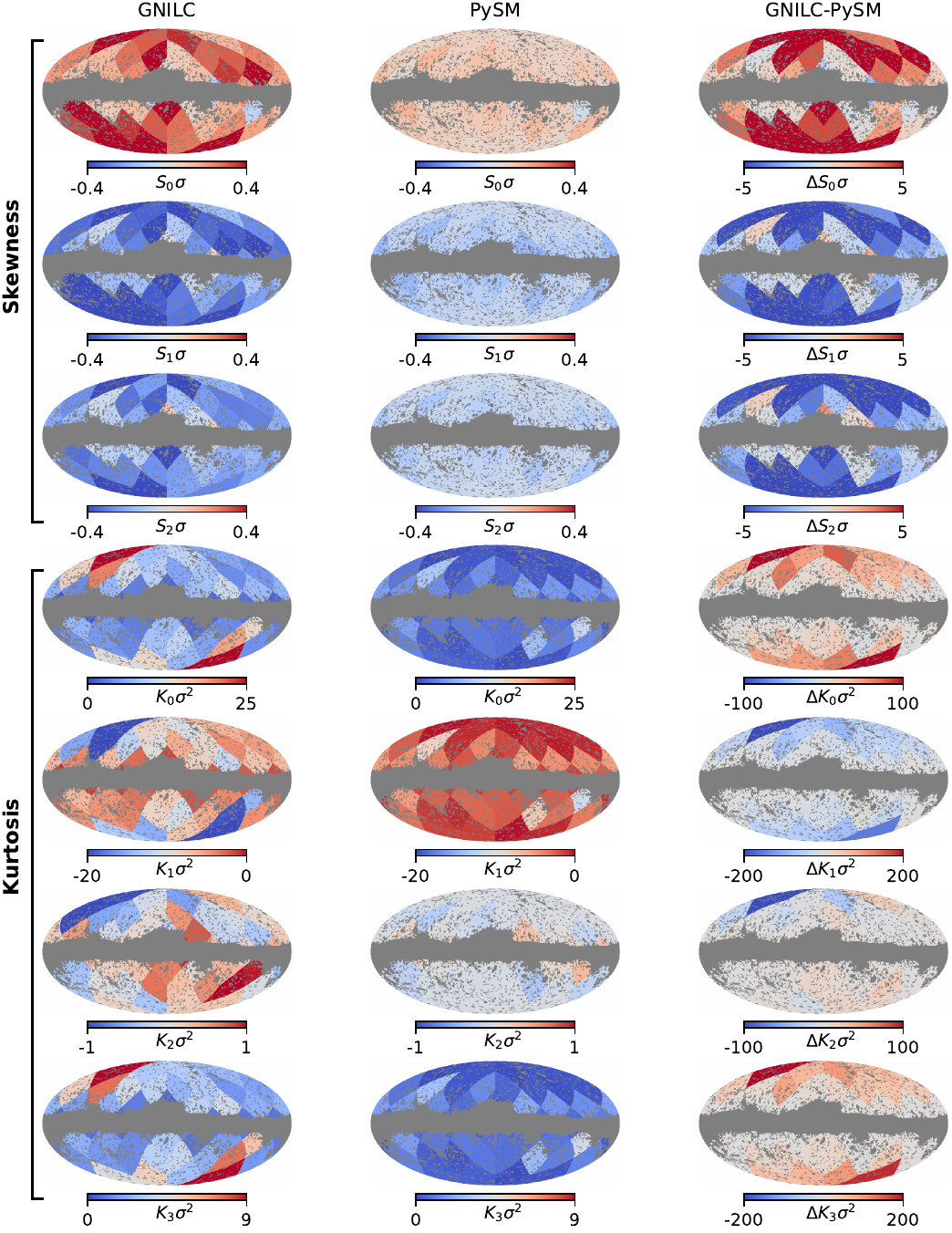}
    \caption{Patch maps of the SK parameters at $\ell_c=300$ for the GNILC dust map (left), the mean over 100 PySM \texttt{d11} realizations (middle), and the significance of their difference computed as $({GNILC}-\langle\mathrm{PySM}\rangle)/\sigma_{\mathrm{PySM}}$ (right). 
    $\sigma_{\mathrm{PySM}}$ denotes the standard deviation estimated from 100 PySM \texttt{d11} realizations.
}
    \label{fig:Skew_patch_maps}
\end{figure}

Next, we perform a patch-based comparison between the GNILC dust map and the PySM model to quantify the spatial variability of non-Gaussianity across the sky. We compare the patch-wise skewness and kurtosis statistics in the data with those in the model realizations. Such localized measurements provide a more stringent test of whether current dust models reproduce the observed spatial distribution of non-Gaussianity in the GNILC map.

We divide the sky into large HEALPix regions corresponding to $N_{\rm side}=2$. This choice balances spatial localization with sufficient sky area to obtain stable estimates of the SK parameters. The SK parameters are then computed within each region of the bandpass-filtered and masked maps described in Section~\ref{sec:DataSets_Preprocessing}, considering only the unmasked pixels within each patch. To avoid sharp boundaries, we apply a $1^\circ$ apodization to the patch edges, similar to the treatment of the analysis mask. Figure~\ref{fig:Skew_patch_maps} shows the patch-wise maps of the SK parameters at $\ell_c = 300$ for the GNILC dust map (left) and the mean PySM signal obtained from 100 \texttt{d11} realizations (middle), together with the significance of their difference, computed as $({\rm GNILC}-\langle{\rm PySM}\rangle)/\sigma_{\rm PySM}$ (right), where $\sigma_{\rm PySM}$ is estimated from the scatter among the PySM realizations. The choice $\ell_c = 300$ lies above the characteristic injection scale of the PySM small-scale realizations ($\ell\sim100$) and well below the multipole cutoff ($\ell^{*}=700$), allowing us to probe small-scale non-Gaussianity while retaining sufficient mode statistics.

For all SK parameters, the GNILC map shows systematically larger amplitudes than the PySM realizations over large regions of the sky, with the discrepancy especially pronounced for the kurtosis parameters. While the GNILC maps exhibit substantial patch-to-patch variation, the corresponding PySM realizations are considerably closer to zero across much of the sky, indicating a suppression of both skewness- and kurtosis-related fluctuations. The significance maps reveal that these differences are spatially widespread rather than being driven by a few isolated regions, indicating a broad mismatch in the higher-order statistical structure of the two maps. These results suggest that, on local sky patches, PySM underestimates the level of small-scale non-Gaussianity present in the GNILC dust map. The fact that the discrepancy appears across all SK parameters further indicates that it extends beyond the one-point PDF to several derivative-based statistics of the field.

\begin{figure}
    \centering
    \includegraphics[width=\columnwidth]{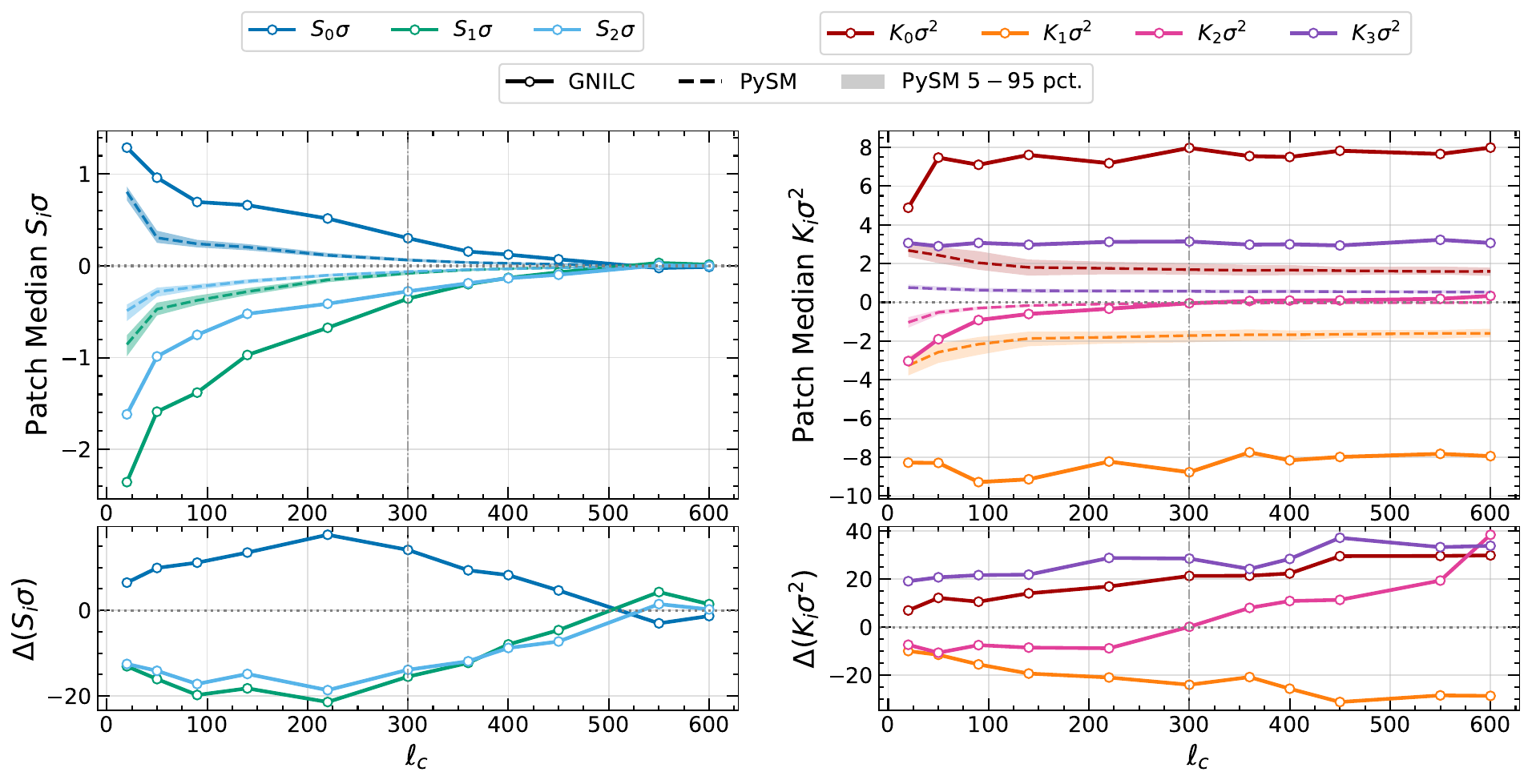}
        \caption{{\it Top}: Patch-wise median skewness (left) and kurtosis (right) parameters as a function of $\ell_c$ for the GNILC dust map (solid) and PySM realizations (dashed). The dashed curves show the median over 100 PySM realizations, with the shaded regions indicating the $5^{\rm th}$--$95^{\rm th}$ percentile range. Vertical dashed lines mark $\ell_c = 300$, corresponding to the scale shown in the patch maps. {\it Bottom}: Difference between the GNILC values and the PySM median, normalized by half of the $5^{\rm th}$--$95^{\rm th}$ percentile range of the PySM realizations.}
    \label{fig:median_ellc}
\end{figure}

To confirm that the map-level differences are not driven by a few extreme regions, we compute the median of the SK parameters over patches with $f_{\rm sky}>0.7$. Figure~\ref{fig:median_ellc} shows the median skewness (left) and kurtosis (right) parameters for GNILC and PySM as functions of scale $\ell_c$. The PySM curves represent the median over realizations, with the shaded band indicating the $5^{\rm th}-95^{\rm th}$ percentile range. For both skewness and kurtosis, the scale dependence closely follows the global trends shown in Figures~\ref{fig:dV_thermdust} and~\ref{fig:dV_dustsims}. However, the patch-median analysis reveals a clearer separation between GNILC and PySM than is apparent in the global measurements, particularly for the kurtosis parameters. Above the injection scale, the PySM realizations remain nearly constant, whereas the GNILC kurtosis parameters remain systematically offset from the PySM values over the same multipole range. Although the GNILC patch maps contain a few high-amplitude regions, this median analysis shows that they are not solely responsible for the GNILC–PySM mismatch. Even after suppressing the contribution from such outliers through the median statistic, the discrepancy between the two maps persists.

These results point to a key limitation in the current PySM small-scale dust construction. Although the model approximately reproduces the scale dependence of the observed MF morphology and SK parameters, both the non-Gaussian amplitudes and the patch-to-patch variations remain lower than in the GNILC map. In the current PySM prescription, most of the spatial inhomogeneity is introduced through the modulation field derived from a smoothed large-scale dust template. However, this modulation is not sufficient to reproduce the level of variation seen in the data. As discussed in Appendix~\ref{secA:pysm_reconstruction}, the dominant small-scale kurtosis originates primarily from the exponential transformation itself, while the modulation procedure contributes only weakly. This points to the need for future dust realizations with more realistic non-Gaussian prescriptions that also capture the observed spatial variation of higher-order statistics. The patch-wise diagnostics introduced here provide a direct way to validate such improvements in foreground modelling.

\section{Discussion on the origin of kurtosis-dominated non-Gaussianity}
\label{sec:OriginofKurt}

The results presented in the previous section reveal a common kurtosis-dominated non-Gaussianity across Galactic foreground maps after filtering out modes below a characteristic multipole $\ell_c$, despite their different astrophysical origins and emission mechanisms. A common feature to note is that the foreground intensities we are studying here are positive definite, and their PDFs generally possess long positive tails. The question then arises whether the small-scale kurtosis-dominance is related to the form of the PDF, and can be explained by simpler statistical considerations. The second aspect of interest is the rate at which the modes approach Gaussianity (i.e., the rate at which $K_i\sigma^{2}$ decreases with increasing $\ell_c$), and whether Gaussianity of modes can be assumed below some scale. This question is tied to the nature of the joint  PDF of the field and its derivatives. 

Qualitatively, let us note that some foreground emission fields on the sky can be roughly modeled in the form
\be
f(\hat n) = \int A(r) g^\gamma(\hat n, r) dr,
\ee
where $r$ is the line-of-sight radial distance, $g$ is some physical random source field, 
$A(r)$ quantifies the non-random spatially dependent factors, and $\gamma$ is some positive-valued index. For example, for synchrotron emissions, $g$ is the underlying Galactic magnetic field, which for small scales may be Gaussian (turbulent part)~\cite{Lazarian2012,Aniket_inprep}. Approximating $A(r)$ as proportional to the Dirac-delta function (which is of course an extreme approximation), the expression  simplifies to
\be
f(\hat n) = G^\gamma(\hat n),  
\ee
where $G$ is a random variable proportional to $g$.  
Then, a foreground field that has more than one random source field may be expressed as
\be
f(\hat n) = G_1^{\gamma_1}(\hat n) +  G_2^{\gamma_2}(\hat n) + \ldots.
\ee
From this approximate but general picture, we may view each foreground field as some linear combination of different powers of $G_i$.  

With the above perspective in mind, we consider three simple classes of random fields constructed from Gaussian `source' fields $G$, but endowed with different one-point PDFs: 
\begin{enumerate}
\item Rayleigh field $f(\hat n) = \sqrt{G^2}$, corresponding to $\gamma=1$ along with the modulus of $G$. 
\item Chi-square field $f(\hat n) = G^2$, with $\gamma=2$. 
\item Lognormal field $f(\hat n) = \exp(G)$, which cannot be associated with a single value of $\gamma$. 
\end{enumerate}
All three fields are strictly positive and positively skewed, but differ in the weight of their high-intensity tails. We generate each field by applying a nonlinear transformation to one or more underlying Gaussian random fields with power-law angular power spectra given by $C_{\ell}\propto \ell^{-\alpha}$. To also probe the role of the underlying spatial correlations, we consider two representative spectral indices, $\alpha=2.5$ and $1.0$, corresponding to $C_{\ell}\propto \ell^{-2.5}$ and $C_{\ell}\propto \ell^{-1}$, respectively. The steeper $\ell^{-2.5}$ spectrum broadly matches the large-scale behaviour of diffuse Galactic foregrounds such as synchrotron and thermal dust emission, while the flatter $\ell^{-1}$ spectrum is included to test the sensitivity of the MF morphology to the underlying spectral slope. The Rayleigh field is constructed from the square root of the sum of squares of two independent Gaussian fields ($f=\sqrt{G_{1}^{2}+G_{2}^{2}}$), the chi-square field from the sum of squares of multiple Gaussian fields ($f=\sum_{i=1}^{k} G_{i}^{2},\, k=5$), and the lognormal field by exponentiating a single Gaussian realization ($f=\exp(-\sigma G),\, \sigma=0.3$). For each case, we generate 1000 full-sky realizations, apply the same bandpass filtering procedure used throughout this work, and compute the MF deviations together with the associated SK parameters.

\begin{figure}[t]
    \centering
    (a)\\    \includegraphics[width=0.95\linewidth]{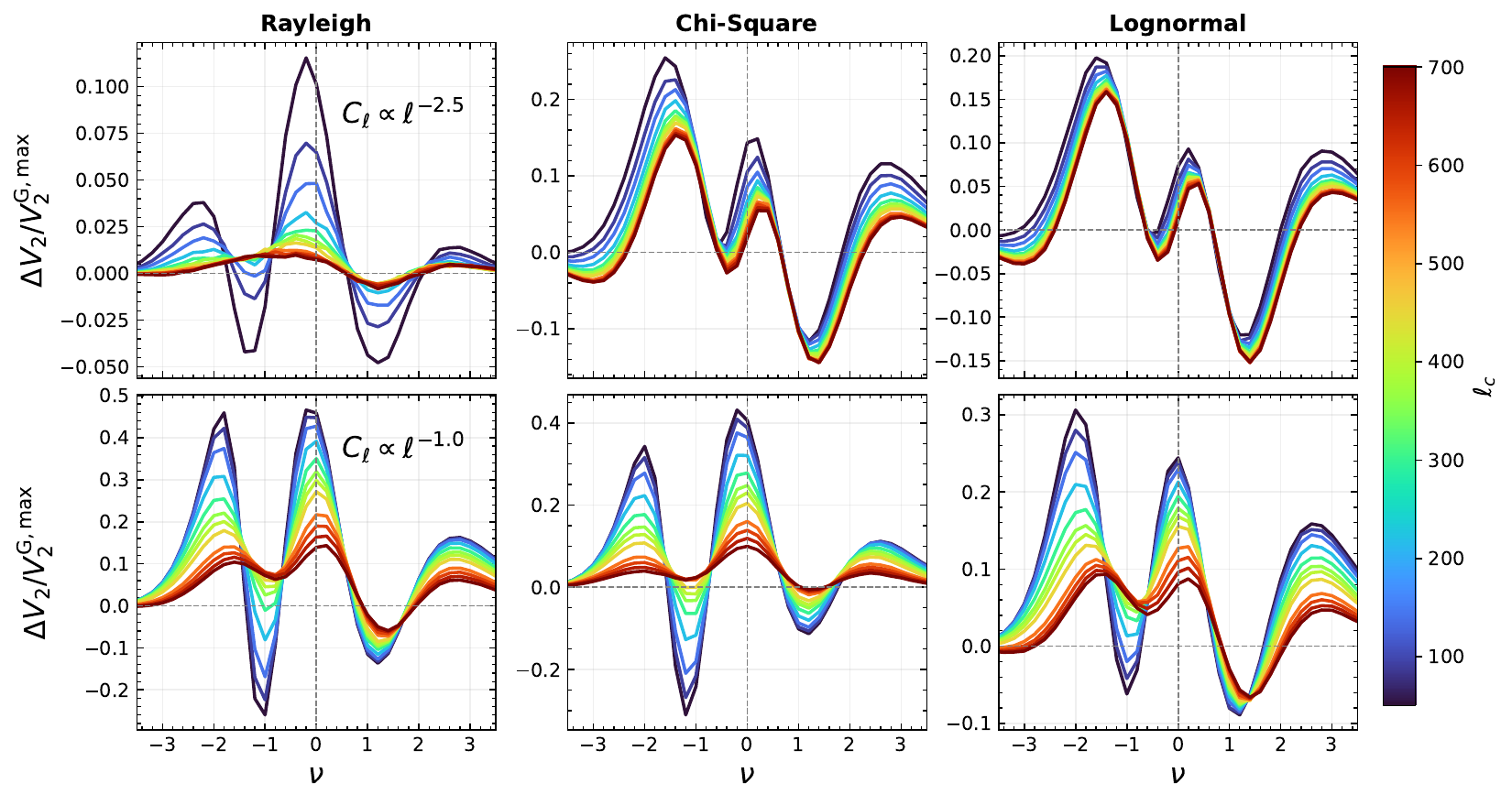}\\
    \vspace{1mm}
  (b)\\  \includegraphics[width=0.98\linewidth]{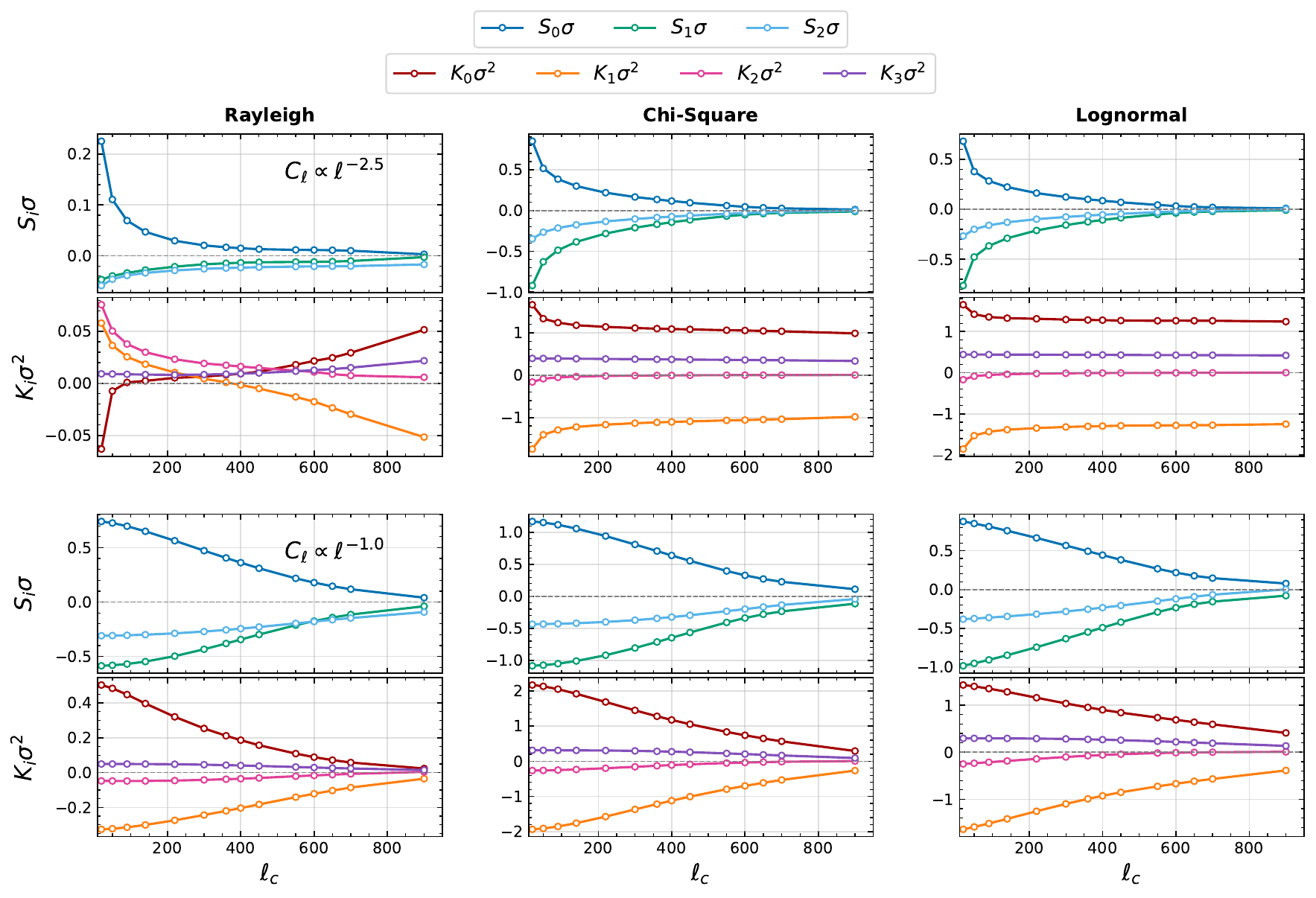}
    \caption{Comparison of the genus non-Gaussian deviations $\Delta V_{2}/V_{2}^{\rm G,max}$ (panel a) and the corresponding generalized skewness and kurtosis parameters (panel b) for simulated non-Gaussian fields with Rayleigh, chi-square, and lognormal one-point PDFs. The top row corresponds to the fiducial red power spectrum $C_\ell \propto \ell^{-2.5}$, while the bottom row shows cases with a flatter spectrum, $C_\ell \propto \ell^{-1.0}$.}
    
    \label{fig:dMF_SK_PDF}
\end{figure}

Figure~\ref{fig:dMF_SK_PDF} (a) shows the genus non-Gaussian deviation ($\Delta V_{2}$) for the three toy models for two different underlying power spectra. For the steeper spectrum, $C_{\ell}\propto \ell^{-2.5}$ (top row), both the chi-square and lognormal fields show a clear kurtosis-dominated non-Gaussian signature at small angular scales, closely resembling the pattern observed in foreground maps. In contrast, the Rayleigh field remains predominantly skewness-dominated and does not develop the kurtosis-dominated morphology seen in the chi-square and lognormal fields. For the flatter spectrum, $C_{\ell}\propto \ell^{-1}$ (bottom row), the MF deviations become more mixed in all three cases and the kurtosis-type structure seen in the steeper-spectrum case is strongly reduced even for the lognormal and chi-square fields. This shows that lognormal- or chi-square–like PDFs alone are not sufficient to reproduce the MF morphology observed in Galactic foreground maps; the underlying spatial correlations of the field also play an important role. Similar trends are also seen in the other MF deviations, $\Delta V_{0}$ and $\Delta V_{1}$. We also tested a wider range of spectral indices and found a systematic transition from mixed skewness–kurtosis MF shapes to a more clearly kurtosis-dominated morphology as the spectrum becomes progressively steeper, with the characteristic foreground-like pattern emerging around $\alpha \gtrsim 2$. Varying the lognormal width parameter $\sigma$ and the chi-square degree parameter $k$ mainly changes the amplitude of the MF deviations, while leaving the overall MF shapes qualitatively unchanged.

To confirm this interpretation, we next examine the scale dependence of SK parameters, shown in Figure~\ref{fig:dMF_SK_PDF} (b). For the steeper spectrum, the skewness parameters rapidly approach zero, while the dominant kurtosis terms remain finite over the full range of scales. As a result, the non-Gaussian signal becomes increasingly kurtosis dominated, similar to that observed in Galactic foreground maps. The Rayleigh field behaves differently even for the same underlying spectrum: for most of the scale range, the skewness parameters exceed the kurtosis contributions, preventing the emergence of a clean kurtosis-dominated regime. This is consistent with the corresponding MF deviations, which exhibit a more skewness-like morphology. For the flatter spectrum, $C_{\ell}\propto \ell^{-1}$ (bottom panel), both the skewness and kurtosis contributions decrease with increasing $\ell_c$, without the emergence of a clean kurtosis-dominated regime, leading to a more mixed skewness–kurtosis signal. These trends match well with the MF results and suggest that the skewness-suppressed, kurtosis-dominated hierarchy observed in Galactic foregrounds is naturally produced by positive-definite fields with pronounced high-intensity tails and sufficiently steep spatial correlations.

Lognormal- and chi-square–like statistics arise naturally from multiplicative line-of-sight processes associated with interstellar turbulence, including density fluctuations, magnetic-field amplification, and the superposition of many emitting regions with fluctuating amplitudes. These processes generate heavy-tailed intensity distributions in which bright compact regions and filamentary features contribute significantly to higher-order moments. At the same time, diffuse Galactic foregrounds exhibit steep spatial correlations over a wide range of scales, often associated with Kolmogorov-like turbulent cascades in the ISM~\cite{Lazarian2012,2024:Stalpes}. To assess whether the toy-model results also apply to Galactic emissions, we repeated the analysis using synthetic dust and synchrotron maps generated from three-dimensional magnetic-field realizations with varying input power-spectrum slopes and projected along the line of sight using simplified emission prescriptions.
We find\footnote{Plots of these results are omitted to avoid clutter.} that these simulations reproduce the trends observed in the toy models, confirming that the latter capture the essential statistical characteristics of diffuse ISM emissions and provide a useful framework for interpreting their higher-order statistical properties.

In summary, using toy PDF models as well as synthetic synchrotron and dust emissions, we have tested two aspects of the statistical nature of foreground emissions, namely, extended high-intensity tails in the one-point PDF and steepness of the correlated structure, to explain non-Gaussianity at small scales. Our results indicate that the combination of these two ingredients naturally gives rise to the common kurtosis-dominated MF signature observed across different Galactic foreground components. 

\section{Summary and conclusions}
\label{sec:DiscussionConclusion}

We present a systematic characterization of non-Gaussianity in Galactic foreground maps using Minkowski Functionals and generalized skewness–kurtosis parameters across different angular scales. For multiple Planck-based tracers, the MF deviations decrease toward smaller angular scales, while the signal becomes increasingly dominated by kurtosis. Although the detailed scale dependence differs between emissions, this kurtosis-dominated trend is seen for all foregrounds despite their diverse astrophysical origins, indicating that it is a robust feature of Galactic emissions.

We then compare these observational trends with commonly used model realizations for thermal dust emission. The PySM lognormal-based dust model reproduces the overall scale dependence of the non-Gaussian statistics, including the approximate scale invariance at small angular scales, but underestimates the kurtosis amplitudes and patch-wise spatial variability observed in the Planck GNILC dust map. The filament-based \texttt{DUSTFILAMENTS} model shows an even weaker non-Gaussian signal, with rapidly decreasing skewness and kurtosis toward smaller angular scales and a noticeably weaker kurtosis-dominated MF morphology. These comparisons show that current dust realizations do not fully reproduce the strength and spatial inhomogeneity of the observed non-Gaussian signal. Improving this will likely require more realistic treatments of higher-order statistics and turbulent small-scale dust structure beyond current prescriptions.

The origin of the observed kurtosis-dominated non-Gaussianity can be understood from the interplay between one-point statistics and spatial structure. Our tests carried out on toy PDF models, as well as synthetic synchrotron and dust emissions, show that heavy-tailed one-point PDFs alone are not sufficient to reproduce the universal MF morphology observed across Galactic foregrounds. The characteristic kurtosis-dominated MF morphology emerges cleanly only when such fields are combined with steep large-scale spatial correlations. In particular, we find a systematic transition from mixed skewness–kurtosis morphology to a strongly kurtosis-dominated regime as the underlying power spectrum becomes progressively steeper, with the foreground-like behaviour appearing around $C_{\ell}\propto \ell^{-2}$ to $\ell^{-2.5}$. This is in line with the diffuse ISM properties, where different radiation fields are known to exhibit both approximately lognormal-like intensity distributions and steep Kolmogorov-like turbulent power spectra. Within this picture, the apparent universality of the kurtosis signal arises from the combination of extended high-intensity tails in the one-point PDF and the correlated large-scale structure associated with interstellar turbulence, while differences in amplitude and scale dependence reflect the distinct physical processes shaping each foreground component.

The results obtained here have direct implications for foreground modelling in CMB analyses. The mismatch between data and current simulations at the level of kurtosis amplitude and spatial variability indicates that small-scale dust structure is not yet adequately captured in existing models. Since residual foreground non-Gaussianity can propagate through component-separation pipelines and impact the recovered CMB maps, this limitation is relevant for higher-order statistics (e.g., primordial non-Gaussianity), lensing reconstruction, and primordial $B$-mode searches. In temperature, direct comparisons between data and simulations are enabled by the availability of high-resolution templates (e.g., GNILC), which allow a direct assessment of small-scale non-Gaussianity. However, in polarization, no equivalent high-resolution observational template currently exists, making it more difficult to validate the high-resolution Q/U model realizations. Given that existing simulation approaches such as PySM adopt a common prescription for temperature and polarization, the limitations identified here may also extend to polarization.

The observed excess kurtosis implies an enhancement of the connected four-point function (trispectrum) of the foreground signal. Quadratic estimators used in CMB lensing reconstruction rely on such four-point statistics and are therefore directly susceptible to bias from foreground non-Gaussianity. Small-scale dust structure can thus induce additional non-Gaussian contributions to the reconstructed lensing signal that are not fully captured by current foreground simulations, motivating dedicated studies to quantify their impact on inferred lensing signals. Existing analyses (e.g., Abril-Cabezas et al.~\cite{Abril-Cabezas:2025whd}) already find biases from non-Gaussian foreground models for upcoming experiments, and if these models underestimate the amplitude of small-scale non-Gaussianity, the true bias may be correspondingly larger.

At the same time, our finding of a universal, kurtosis-dominated non-Gaussian morphology across different foreground components provides a useful guide for foreground modelling and a direct benchmark for the construction of small-scale non-Gaussian structure in simulations, complementing component-specific prescriptions. A natural next step is therefore to develop simulations that reproduce the observed signal more realistically by incorporating controlled non-Gaussian statistics, as explored in previous studies~\cite{Huffenberger:2004gm,MivilleDeschenes:dustCl,Martire:2023ytg}. Extending existing frameworks such as PySM to include information from higher-order statistics, along with realistic spatial modulation, would help achieve a better description of the small-scale foreground statistics across the sky. Emerging approaches based on machine learning and scattering transforms offer promising avenues. Physics-informed models such as \texttt{DUSTFILAMENTS}, which are well suited for understanding dust-induced biases in cosmic birefringence measurements~\cite{Hervias-Caimapo:2024ili}, will require further development to tune the filament properties to match the observed level of skewness and kurtosis. MHD simulations of the interstellar medium will also play an important role in studying how turbulent properties such as sonic and Alfv\'enic Mach numbers shape the bispectrum, trispectrum, and MF signatures of Galactic foregrounds, helping connect our observations to the underlying ISM physics~\cite{Hu:2019bvz,Saydjari:2021,Wang:2024tku,Williamson:2024,2024:Stalpes}.

More generally, the statistical framework we present here provides a direct way to validate improvements in foreground modelling at the map level, alongside standard power-spectrum-based validation. Beyond foreground modelling, these MF-based tools serve as general diagnostics for quality assessment in CMB experiments, helping identify residual contamination, leakage, and systematic effects in reconstructed maps. They can reveal subtle morphological differences that remain hidden in conventional analyses, providing a useful complement to existing consistency checks.

\acknowledgments 
{FR acknowledges the use of the \texttt{Memphis} and \texttt{Grace} HPRC clusters at Texas A\&M University and the \texttt{Nova} cluster at the Indian Institute of Astrophysics, Bengaluru. FR and KH are supported by NASA award 80NSSC25K7506. FR thanks Jacques Delabrouille, Shamik Ghosh, Brandon Hensley, and Blake Sherwin for useful discussions and Carlos Herv\'ias-Caimapo for providing the \texttt{DUSTFILAMENTS} realizations used in this work. FR also thanks Will Coulton for hospitality at the Kavli Institute for Cosmology, Cambridge, where part of this work was carried out. The Planck products used in this work were obtained from the Planck Legacy Archive (PLA). 
This work made use of 
\texttt{healpy}~\cite{Zonca:healpy}, 
\texttt{PySM}~\cite{Thorne2017}, 
and \texttt{NaMaster}~\cite{NaMaster:2019} packages.}

\appendix

\section{Perturbative expansion of MFs for weakly non-Gaussian fields}
\label{SecA:weaklynGMF}

For a mildly non-Gaussian isotropic field with zero mean, the ensemble expectations of the MFs are given by~\cite{Matsubara:2011,Gay:2012,Adler:1981,Tomita:1986}
\begin{equation}
V_{k}(\nu)=A_k\,e^{-\nu^{2}/2}v_k(\nu), \quad A_{k}=\frac{1}{(2\pi)^{(k+1)/2}}\frac{\omega_{2}}{\omega_{2-k}\omega_{k}}\Big(\frac{\sigma_{1}}{\sqrt{2}\sigma}\Big)^{k},
\label{eqn:gmf}
\end{equation}
where $\sigma_{1}$ denotes the standard deviation of the field gradient, $k=0,1,2$, and $\omega_0=1,\ \omega_1=2,\ \omega_2= \pi$. 
The functions $v_k$ are given as perturbative expansions in powers of $\sigma$, 
\begin{equation}
v_{k}=v_{k}^{(\rm G)}+v_{k}^{(1)}\sigma+v_{k}^{(2)}\sigma^{{2}}+\mathcal{O}(\sigma^{3}). 
\end{equation}  
The zeroth order terms correspond to the expressions for Gaussian fields, given by
\begin{eqnarray}
 v_0^{\rm (G)}(\nu) = \sqrt{\frac{\pi} {2}}\,e^{\nu^{2}/2}\,\text{erfc}\left(\frac{\nu}{\sqrt{2}}\right), \quad 
v^{(\rm G)}_1 = 1, \quad
v^{(\rm G)}_2(\nu) = \nu.
\end{eqnarray}
The first-order corrections are determined by three generalized skewness parameters $S_0, S_1, S_2$, with $v_k^{(1)}$ given by
\begin{eqnarray}
v_{0}^{(1)}(\nu)&=&\frac{S_0}{6}H_{2}(\nu),\\
v_{1}^{(1)}(\nu)&=&\frac{S_0}{6}H_{3}(\nu)-\frac{S_1}{4}H_{1}(\nu),\\
v_{2}^{(1)}(\nu)&=&\frac{S_0}{6}H_{4}(\nu)-\frac{S_1}{2}H_{2}(\nu)-\frac{S_2}{2}H_{0}(\nu),
\end{eqnarray}
where $H_n(\nu)$ denotes probabilists' Hermite polynomials. 
The second-order terms depend on the four generalized kurtosis parameters $K_0, K_1, K_2, K_3$,
\begin{eqnarray}
  v_{0}^{(2)}(\nu)&=&\frac{S_0^2}{72}H_{5}(\nu)+\frac{K_0}{24}H_{3}(\nu), \\
v_{1}^{(2)}(\nu)&=&\frac{S_0^2}{72}H_{6}(\nu) +\frac{K_0-S_0S_1}{24}H_4(\nu)
   -\frac{1}{12}  \left(K_1+\frac{3}{8}S_1^2\right)  H_2(\nu)-\frac{K_3}{8}, \\
   v_{2}^{(2)}(\nu)&=&\frac{S_0^2}{72}H_{7}(\nu)+\frac{K_0-2S_0S_1}{24}H_{5}(\nu)-\frac{1}{6}\left(K_1+\frac{1}{2}S_0S_2\right)H_{3}(\nu) 
    -\frac{1}{2}\left(K_2+\frac{1}{2}S_1S_2\right)H_{1}(\nu). \nn\\ 
\end{eqnarray}
While the above expressions are formally derived for weakly non-Gaussian fields, they nevertheless provide useful intuition for interpreting how the generalized skewness and kurtosis parameters shape the MF deviations discussed in the main text.

\section{\textit{Commander} dust and Planck 545 GHz maps}
\label{secA:commander_545}

\begin{figure}
    \centering    \includegraphics[width=1.0\linewidth]{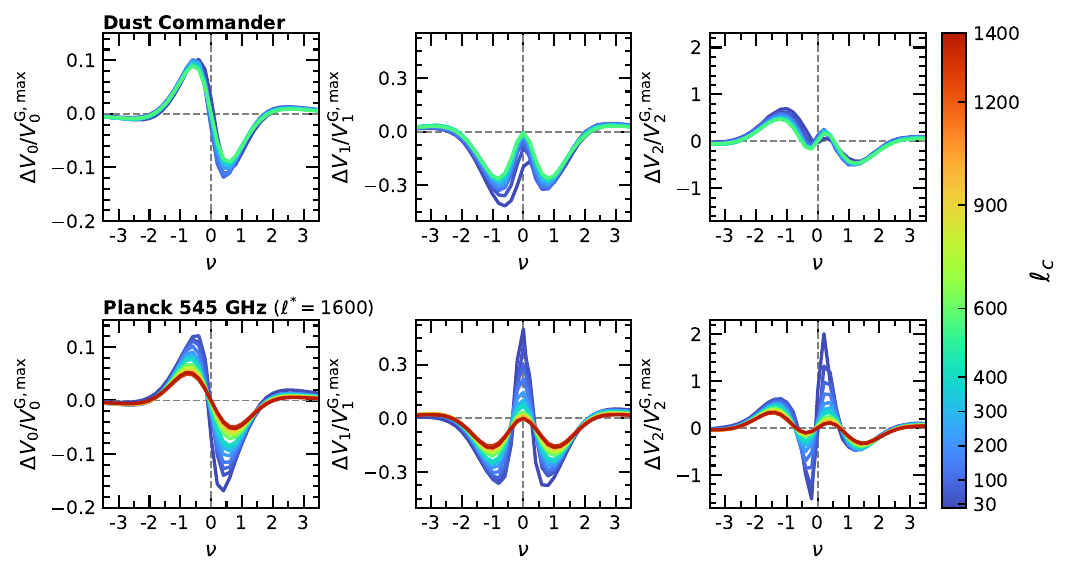}
    \\
    \includegraphics[width=0.92\linewidth]{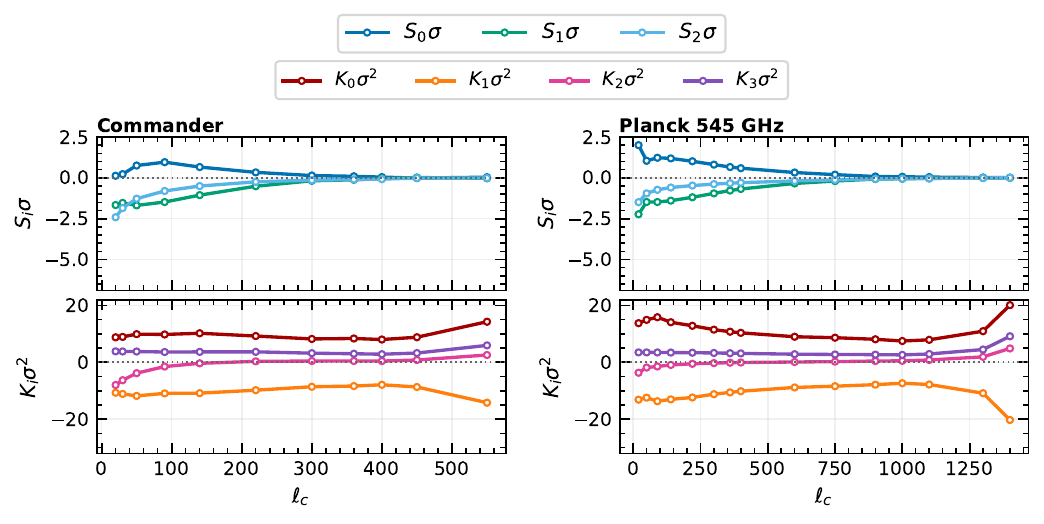} 
    \caption{MF deviations (top panels) and SK parameters (bottom panels) for the Planck \textit{Commander} dust map and the Planck 545\,GHz map. The colors denote the filtering scale $\ell_c$. The 545\,GHz analysis is extended to higher multipoles, up to $\ell^\ast=1600$.}
    \label{fig:SK_MF_commander_545}
\end{figure}

In this section, we present the results for the high-resolution Planck \textit{Commander} dust temperature map as an alternative component-separation product~\cite{Planck_2015_X}. We also extend the analysis of the 545\,GHz frequency map to higher multipoles by adopting a larger upper cutoff, $\ell^{*}=1600$, taking advantage of its native angular resolution of approximately 5 arcmin. This enables us to probe the non-Gaussian behaviour over a wider multipole range. The corresponding SK parameters and MF deviations are shown in Figure~\ref{fig:SK_MF_commander_545}.

The \textit{Commander} dust map shows SK and MF trends broadly similar to those seen in the other Planck dust tracers. The kurtosis terms dominate over the skewness contributions and remain nearly constant at small angular scales, consistent with the behaviour seen in the GNILC and 545\,GHz maps. The MF deviations likewise retain the characteristic kurtosis-dominated morphology. However, the overall amplitudes of the SK parameters are lower, and the MF deviations are correspondingly reduced compared to the GNILC and 545\,GHz results. This difference likely reflects the distinct component-separation methodology, as well as differing levels of residual contamination from the Cosmic Infrared Background (CIB) and instrumental noise.

At higher resolution, extending beyond the $\ell^\ast=700$ limit used in the main text, the Planck 545\,GHz map shows that the kurtosis-dominated plateau in the SK parameters persists to smaller angular scales. The mild increase previously seen around $\ell_c \sim 600$ shifts toward values closer to the upper cutoff scale $\ell^\ast$, indicating that it is likely related to the decreasing number of modes near the edge of the bandpass filter rather than a physical feature. A similar trend is also visible, to varying degrees, in the GNILC and \textit{Commander} dust maps. The MF deviations continue to decrease gradually toward smaller scales, indicating a suppression of the overall level of non-Gaussianity. It should also be noted that these scales correspond to regimes where Galactic dust becomes less dominant, with increasing contributions from the CIB, instrumental noise, and other small-scale contaminants.

\section{Origin of non-Gaussianity in the PySM reconstruction}
\label{secA:pysm_reconstruction}

To better understand the origin of the non-Gaussian signatures in the PySM dust maps, we analyze the one-point skewness and kurtosis amplitudes, $S_0 \sigma$ and $K_0 \sigma^2$, at different stages of the PySM small-scale reconstruction~\cite{PySM2025}. To construct the PySM \texttt{d9} model, the GNILC dust map is first transformed into log-space. The power spectrum of this log-space map is then extrapolated to smaller angular scales to generate Gaussian small-scale realizations. A smooth large-scale template map is then applied to modulate the amplitude of the small-scale Gaussian fluctuations according to the underlying dust morphology, after which the resulting realization is added to the low-pass filtered large-scale dust component. The final high-resolution dust map is obtained by mapping the field back from log-space using an exponential transformation.

\begin{figure}[t]
    \centering
    \includegraphics[width=0.99\linewidth]{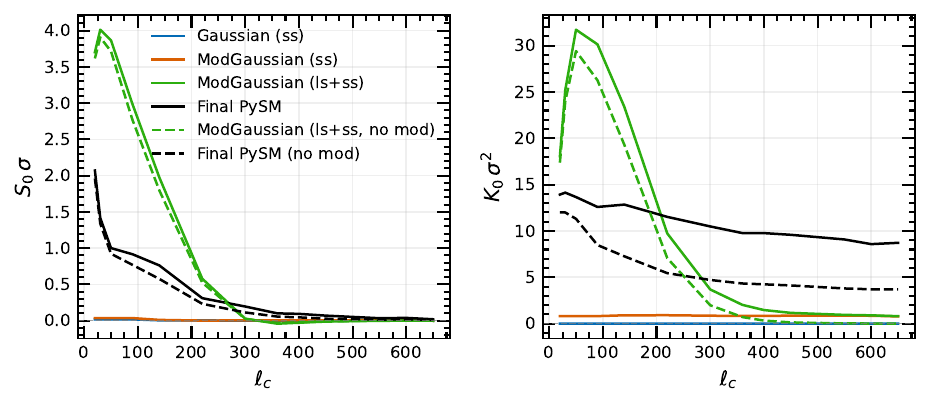}
    \caption{Skewness $S_0\sigma$ and kurtosis $K_0\sigma^{2}$ as a function of bandpass scale $\ell_c$ for different stages of the PySM reconstruction. Solid curves correspond to the standard PySM pipeline, while dashed curves show the corresponding unmodulated realizations.}
    \label{fig:pysm_dust_construction}
\end{figure}

Using the intermediate products provided by the PySM team, we separately analyze (a) the Gaussian small-scale realization, (b) the modulated Gaussian small-scale realization, (c) the realization after adding the large-scale dust template, and (d) the final map after transforming back from log-space. For stages (c) and (d), we also analyze corresponding realizations constructed without applying the modulation map, shown by the dashed curves. This allows us to isolate the contribution of each stage of the PySM construction and assess the role of the modulation step in generating the final non-Gaussian signal. The results are shown in Figure~\ref{fig:pysm_dust_construction}.

We first examine the skewness amplitude $S_0 \sigma$. As expected, the purely Gaussian realization exhibits vanishing skewness across all bandpass scales. Introducing multiplicative modulation alone produces only a negligible skewness signal. However, after adding the large-scale dust template, the skewness becomes strongly enhanced at low $\ell_c$, reflecting the intrinsically non-Gaussian large-scale structure of the GNILC dust field. The contribution from the large-scale dust template rapidly decreases toward smaller angular scales, where the Gaussian small-scale realization becomes dominant. The final lognormal mapping then produces a persistent skewness signal that remains significant even at high $\ell_c$. The exponential transformation stretches positive fluctuations more strongly than negative ones, generating an asymmetric one-point distribution. The skewness of the final PySM template at high $\ell_c$ is therefore set primarily by the lognormal transformation, while the large-scale dust template mainly controls the excess skewness at low $\ell_c$.

Turning to the kurtosis amplitude $K_0 \sigma^2$, the purely Gaussian realization again exhibits vanishing signal at all $\ell_c$. Introducing multiplicative modulation generates a small but non-zero kurtosis signal that remains nearly constant across scales. After adding the large-scale dust template, the kurtosis becomes strongly enhanced at low $\ell_c$, reflecting the non-Gaussian large-scale structure of the GNILC dust field, before asymptotically approaching the modulated small-scale realization at high $\ell_c$. Finally, the lognormal transformation produces a much stronger, approximately scale-independent plateau in $K_0 \sigma^2$ at small angular scales, showing that the nonlinear mapping is the primary source of the persistent kurtosis in the final PySM template. This behaviour is also consistent with the toy-model results discussed in the main text, where lognormal fields constructed from steep dust-like power spectra naturally develop a stable kurtosis-dominated regime at high $\ell_c$. Since the PySM small-scale dust realization is generated from extrapolated dust power spectra together with this nonlinear mapping, the high-$\ell_c$ kurtosis plateau can be understood as a direct consequence of the same mechanism.

The comparison between the modulated and unmodulated realizations further clarifies the role of the modulation step in the PySM construction. Multiplicative modulation produces only a negligible skewness signal because it mainly rescales the local variance of the Gaussian field without introducing a preferred sign. The one-point distribution therefore remains approximately symmetric even after modulation. At the same time, the kurtosis acquires a small but nearly scale-independent excess after modulation. This arises because the modulation map generates a mixture of regions with different local variances, broadening the one-point distribution and increasing the higher-order even moments. This behaviour is evident from the comparison between the solid and dashed curves in Figure~\ref{fig:pysm_dust_construction}: modulation has only a weak effect on the skewness amplitudes, while producing a noticeable increase in the kurtosis amplitudes across scales. The relatively weak impact of modulation on the SK amplitudes suggests that, while it introduces spatial inhomogeneity in the small-scale realization, it may not be sufficient to reproduce the observed patch-to-patch variation of the non-Gaussian signal. This points to the need for more realistic treatments of the modulation field.

\section{Comparison of GNILC and PySM maps and one-point PDFs}

\begin{figure}[t]
    \centering
    \includegraphics[width=0.88\linewidth]{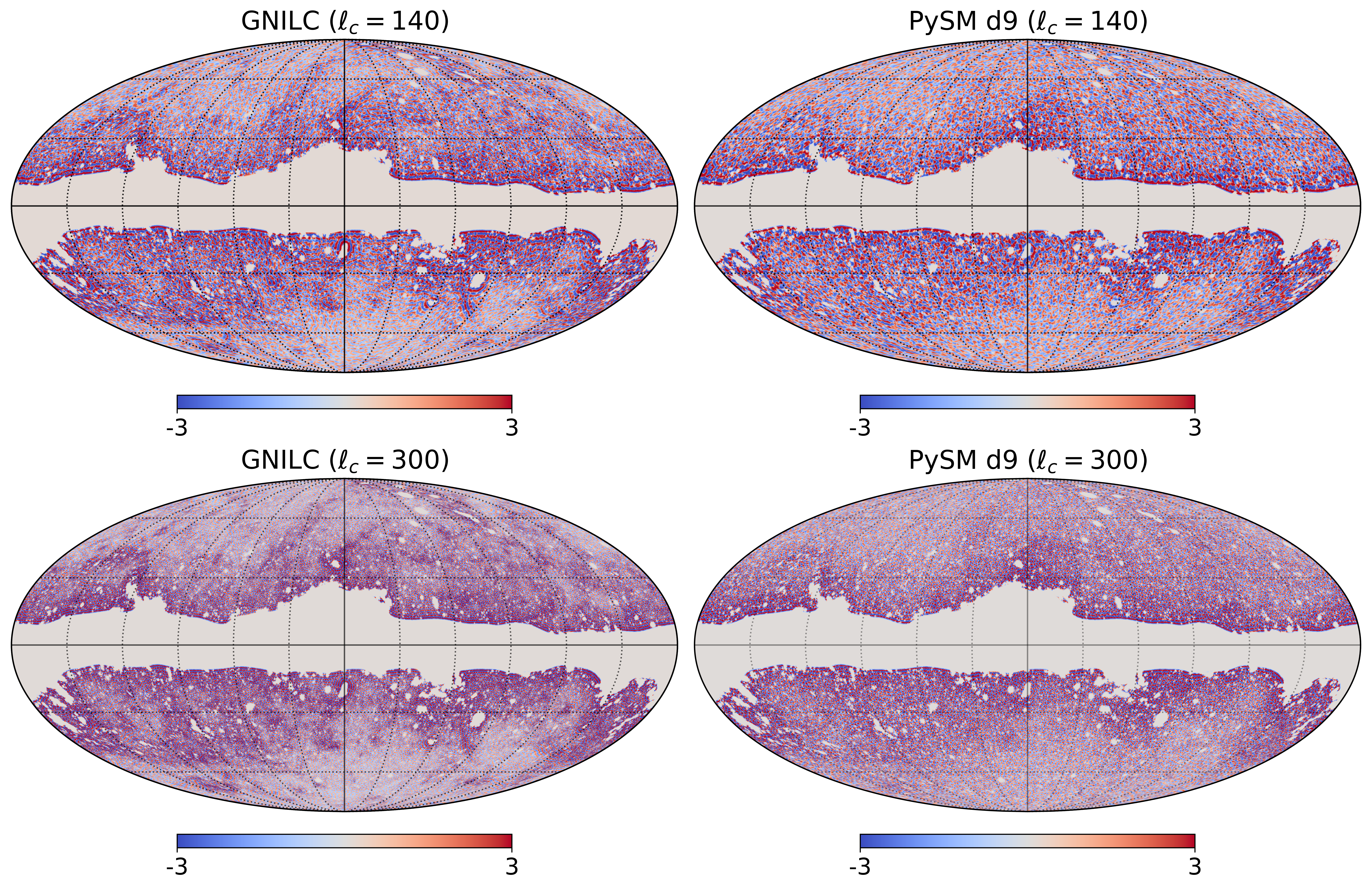} \\
    \includegraphics[width=0.88\linewidth]{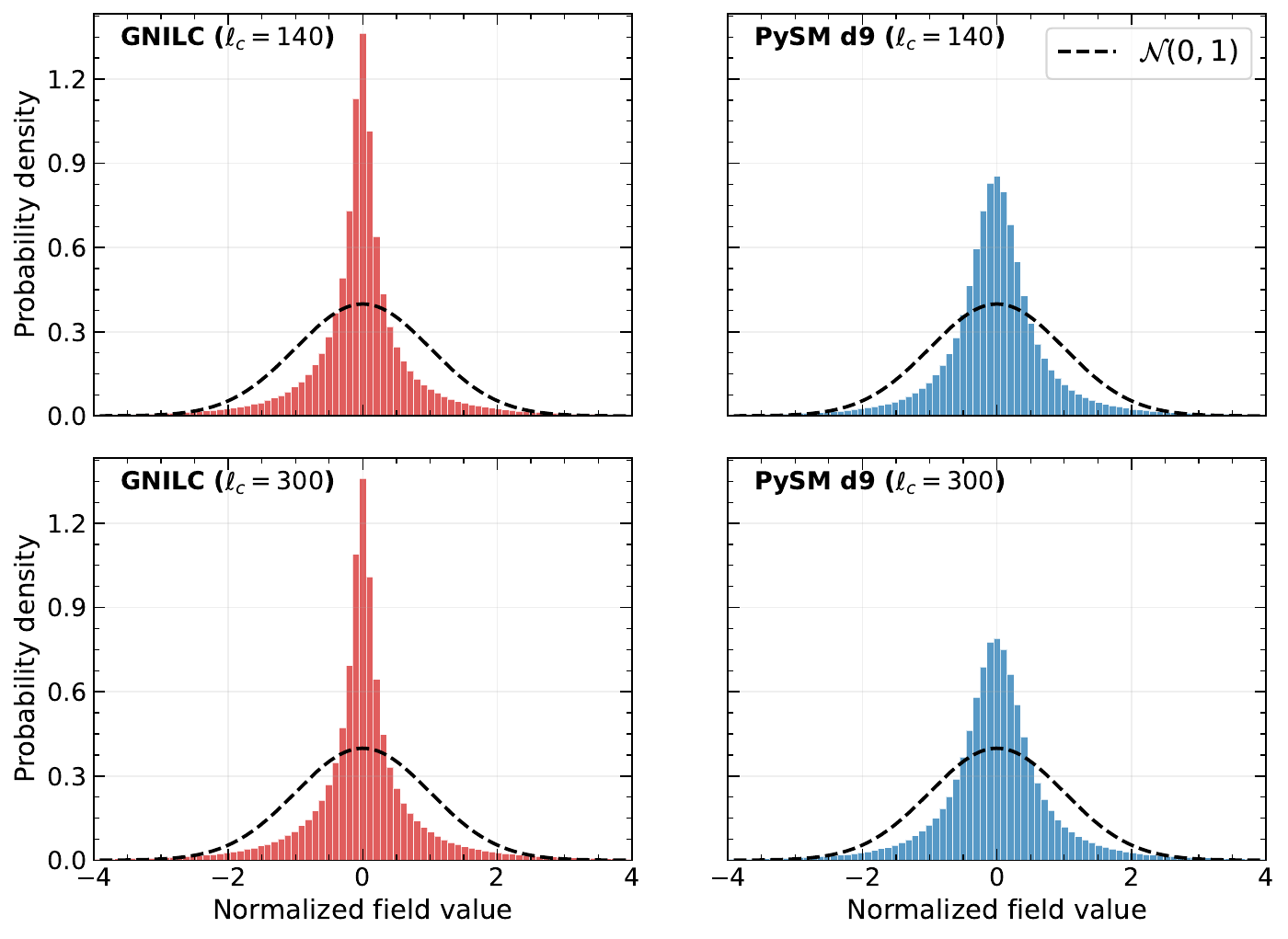}
    \caption{Comparison between the GNILC and PySM thermal dust maps after bandpass filtering at two angular scales, $\ell_c = 140$ and $\ell_c = 300$, shown after mean subtraction and variance normalization. The upper panels show the filtered maps using the same color scale, while the lower panels show the corresponding one-point probability density functions compared with a unit-variance Gaussian distribution. The GNILC maps exhibit a sharper central peak and mildly broader non-Gaussian tails than PySM \texttt{d9}, consistent with the larger kurtosis amplitudes inferred from the MF and SK analyses.}
    \label{fig:gnilc_pysm_map_hist}
\end{figure}

Figure~\ref{fig:gnilc_pysm_map_hist} compares the small-scale thermal dust structures in the {\it GNILC} and PySM \texttt{d9} maps after bandpass filtering at $\ell_c = 140$ and $\ell_c = 300$. The maps are shown after mean subtraction and variance normalization using a common color scale, while the lower panels show the corresponding one-point probability density functions compared with a unit-variance Gaussian distribution. Although both maps exhibit non-Gaussian small-scale structure, the {\it GNILC} map shows a sharper central peak and mildly broader non-Gaussian tails in its one-point distribution than the PySM map. These visual differences are consistent with the trends seen in the MF and SK analyses presented above.

\section{MF deviations in area-fraction threshold values}
\label{secA:pixel_shuffle_nuA}

To isolate the contribution of spatial morphology from the dominant one-point PDF contribution, we additionally examine the MF deviations using the area-fraction threshold parametrization. In this representation, the thresholds are expressed in terms of a Gaussian-equivalent variable, denoted by $\nu_A$. Instead of the usual normalized field threshold $\nu$, the parametrization is defined through the excursion-set area fraction,
\begin{equation}
V_0=\frac{1}{\sqrt{2\pi}}\int_{\nu_A}^{\infty} e^{-t^2/2}\,dt,
\end{equation}
such that $\nu_A$ corresponds to the threshold of a Gaussian field having the same area fraction. This transformation largely removes the direct contribution of the one-point PDF from the threshold parametrization while preserving the spatial structure and phase correlations of the field. MF deviations expressed as a function of $\nu_A$ therefore provide a useful way to separate the effects of the one-point PDF from those of the underlying spatial correlations.

\begin{figure}[t]
    \centering    \includegraphics[width=0.99\linewidth]{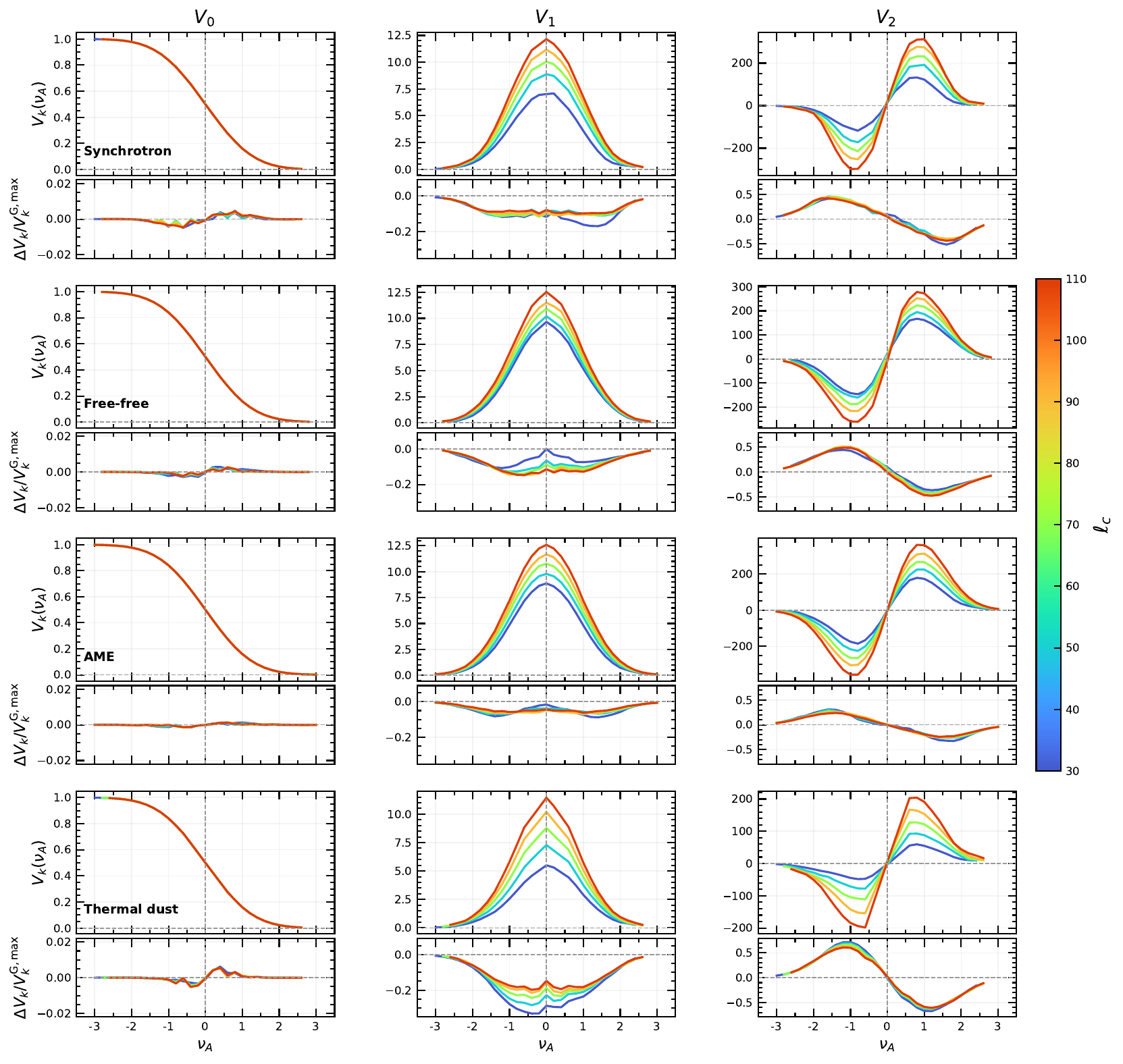} 
    \caption{Comparison of the Minkowski functionals $V_k(\nu_A)$ (upper row of each component block) and the corresponding non-Gaussian deviations $\Delta V_k/V_k^{G,\max}$ (lower row) for the Planck Commander foreground components in the area-fraction threshold parametrization $\nu_A$. From top to bottom, the panels correspond to synchrotron, free-free, AME, and thermal dust emission. Different colors denote different bandpass filtering scales $\ell_c$.}
    \label{fig:fg_comm256_nuA}
\end{figure}

Figure~\ref{fig:fg_comm256_nuA} shows the MFs $V_k$ and the corresponding MF deviations $\Delta V_{k}$ as a function of $\nu_A$ for the four \textit{Commander} foreground components: thermal dust, synchrotron, free-free, and AME. In the ordinary $\nu$ parametrization discussed in the main text, all foregrounds exhibit broadly similar kurtosis-dominated MF shapes despite their different spatial structures. After transforming to the $\nu_A$ representation, this common kurtosis-dominated contribution is substantially reduced. The MF deviations now show much stronger component dependence, especially in $V_1$, where the amplitudes, peak locations, and overall threshold dependence differ significantly between the foregrounds. This observation is consistent with the interpretation developed in the main text: the combination of an approximately lognormal-like one-point PDF and steep spatial correlations produces the common kurtosis-dominated MF morphology, while the remaining differences reflect the distinct spatial structure of each foreground component.

\def\apj{ApJ}%
\def\mnras{MNRAS}%
\def\aap{A\&A}%
\def\apjl{ApJ}
\def\aj{AJ}
\def\physrep{PhR}
\def\apjs{ApJS}
\def\jcap{JCAP}
\def\pasa{PASA}
\def\pasj{PASJ}
\def\nat{Natur}
\def\apss{Ap\&SS}
\def\araa{ARA\&A}
\def\aaps{A\&AS}
\def\ssr{Space Sci. Rev.}
\def\pasp{PASP}
\def\na{New A}
\def\prd{PRD}

\bibliography{fgref}
\bibliographystyle{JHEP}

\end{document}